\documentclass[fleqn,usenatbib]{mnras}

\usepackage{newtxtext,newtxmath}
\usepackage[normalem]{ulem}
\usepackage{soul}
\usepackage[T1]{fontenc}

\DeclareRobustCommand{\VAN}[3]{#2}
\let\VANthebibliography\thebibliography
\def\thebibliography{\DeclareRobustCommand{\VAN}[3]{##3}\VANthebibliography}

\usepackage{graphicx}	% Including figure files
\usepackage{amsmath}	% Advanced maths commands
\usepackage{adjustbox}

\title[The cosmic journey of dust]{The cosmic journey of dust grains -- from nucleation to planetary system}

\author[K. Lund et al.]{
Kira Lund,$^{1}$\thanks{E-mail: lund-k@outlook.com}
Anders Johansen,$^{1,2}$ and
Oscar Agertz$^{1}$
\\
$^{1}$Lund Observatory, Division of Astrophysics, Department of Physics, Lund University, Box 43, SE-221 00, Sweden\\
$^{2}$Centre for Star and Planet Formation, Globe Institute, University of Copenhagen, Øster Voldgade 5-7, 1350 Copenhagen, Denmark\\
}

\date{Accepted XXX. Received YYY; in original form ZZZ}

\pubyear{\the\year{}}

\begin{document}
\label{firstpage}
\pagerange{\pageref{firstpage}--\pageref{lastpage}}
\maketitle

\begin{abstract}
 Dust is essential to the evolution of galaxies and drives the formation of planetary systems. The challenge of inferring the origin of different presolar dust grains from meteoritic samples motivates forward modelling to understand the contributions of low- and high-mass stars to dust in our Solar System. In this work we follow the evolution of dust with tracer particles within a hydrodynamical simulation of a Milky Way-like isolated disc galaxy. We find that nearly half of the grains released from stars lose less than $10\%$ of their initial mass due to thermal sputtering in the interstellar medium (ISM), with an average degree of atomisation $\sim$10\% higher for dust grains released by supernovae relative to asymptotic giant branch (AGB) star grains. We show through supernova remnant model variations that supernova (SN) dust survival is primarily shaped by the supernova bubble environment in the first million years ($\mathrm{Myr}$) after the explosion rather than by its evolution during $10^2-10^3\,\mathrm{Myr}$ in the ISM. The AGB/SN ratio of dust grains incorporated into newly formed stars approaches $0.8$ after a few hundred $\mathrm{Myr}$ of galactic evolution. Our analysis also shows that star-forming particles with short ($<$$10 \, \mathrm{Myr}$) free--floating time-scales in the ISM are predominantly released from supernovae rather than AGB stars. This implies that the Solar System budget of short-lived radioactive isotopes such as $^{26}$Al, whose decay contributed to melting and differentiating planetesimals, should have been provided by massive stars with masses $M \gtrsim 8 \, \text{M}_{\sun}$.

\end{abstract}


%%%%%%%%%%%%%%%%%%%%%%%%%%%%%%%%%%%%%%%%%%%%%%%%%%

%%%%%%%%%%%%%%%%% BODY OF PAPER %%%%%%%%%%%%%%%%%% 

\section{Introduction}
\label{sec:intro}
Dust contributes to at most a few per cent of the baryonic mass of galaxies \citep[e.g.,][]{Shull_2012, Ruyer_2014}, yet dust plays an essential role in star formation and galactic evolution. Much of our understanding of galaxies comes from their spectral energy distribution \citep[e.g.][]{Tully_1977, Kennicutt_1998, Calzetti_2000}. Light in the optical and ultraviolet range is efficiently absorbed by dust and subsequently re-emitted in the infrared \citep[for a review see][]{Draine_2003}. Consequently, the spectral energy distributions of galaxies are strongly shaped by the properties of dust grains, particularly their size and chemical composition. Dust is therefore a major source of opacity in the Universe and can also serve as an important tracer of cold gas. \citep[see e.g.,][]{Eales_2012, Groves_2015, Schinnerer_2016}. Apart from its observational relevance, dust regulates the chemistry and the thermal balance of the interstellar medium \citep[][]{Bakes_1994, Hocuk_2014}, and consequently the formation of stars and planets. Grain surfaces also catalyse chemical reactions critical to the formation of new stars, such as molecular hydrogen \citep[$\text{H}_2$;][]{Hollenbach_1971}.

The evolution of dust in a galactic context has been studied primarily in one- or two-zone chemical evolution models in the past \citep[e.g.,][]{Dwek_1998, Hirashita_2002, Morgan_2003, Gall_2011, Zhukovska_2014, Feldmann_2015}. These models assess the characteristics of dust averaged over the entire galaxy, or larger regions of it, but do not capture dust processing on smaller scales. However, observed variations in element depletions in the interstellar medium \citep{Jenkins_2009}, for instance, indicate that grain properties strongly depend on their local environment. This implies that a three-dimensional approach is required to understand and correctly model the evolution of dust. More recent efforts to incorporate and study the effects of dust in isolated galaxy or cosmological hydrodynamical simulations have been made \citep[e.g.,][]{Bekki_2013, Bekki_2015, McKinnon2016, Aoyama_2016, Mckinnon_2017, Hou_2019, Vogelsberger_2019, Graziani_2020, Granato_2021, Dubois_2024}. Due to the computational cost of considering the entire grain size distribution, most of these works employ the two-size approximation proposed by \cite{Hirashita_2015}, separating particles into small and large grains below and above $0.03 \, \micron$, respectively. While this simplification may provide a means of investigating dust evolution in a cosmological or galactic setting at reduced computation time, it is evidently unable to represent variations in grain size distribution. More importantly, however, large-scale simulations are strongly dependent on subgrid models that bring about parameter degeneracies and their associated uncertainties \citep[see e.g.,][]{Lower_2024}. An alternative approach implemented by e.g. \citet{Esmerian_2022} is to investigate the evolution of dust in post-processing. Despite not incorporating the effects of dust on the chemistry and thermodynamics of the ISM explicitly, this kind of analysis allows for the exploration of a wide range of models at minimal computational effort.               

Observations of galaxies of redshift $z \geq 6$ that contain substantial amounts of dust, presented in \citet{bertoldi_dust_2003, Beelen_2006, Watson_2015, Wang_2021}, suggest a rapid dust production scenario due to short-lived massive stars. Core--collapse supernovae, more specifically SNe type II are expected to be important contributors to the dust inventory of galaxies \citep[e.g.][]{Szalai_2013, Sarangi_2015}, yet supernova shock waves have long been considered as the primary mechanism by which grains are destroyed both in the supernova remnant (SNR) and in the ISM \citep[for a review on dust processing in the early Universe see][]{Schneider_2024}. However, no consensus has been reached on quantifying the disruptive effect of forward shocks on grains in the circumstellar medium, nor that of the reverse shock wave on dust in the SN ejecta. \cite{Nozawa_2007} determine a survival fraction between 0 and $80\%$ of the dust mass initially produced before being injected into interstellar space, depending on the explosion energy and the density of the surrounding ISM. \cite{Bianchi_2007}, on the other hand, find that only 2--$20 \%$ of the original dust mass remains for varying ISM densities. This inconsistency is only further compounded by the variability in survival fractions reported in works that consider different models for the dust evolution, as well as physical conditions in the ISM, leading to dust survival in the range of e.g., 0--$99\%$ (\citealp[][]{Silvia_2010, Silvia_2012}\citealp[, but also][]{Kirchschlager_2023}), 1--$8\%$ in \cite{Bocchio_2016}, 12--$16\%$ in \cite{Micelotta_2016}, 0--$40\%$ or 17--$28\%$ \citep{Kirchschlager_2019, Kirchschlager_2024b}. Several of these studies assess the processing of dust in a homogeneous medium, yet the results of \citet{Kirchschlager_2024a} indicate that the inhomogeneity of the ISM, as well as turbulence and magnetic effects may decrease dust destruction up to a factor of 2. Additional support in this direction comes from observational evidence by \citet{Priestley_2001}, whose fits to the infrared emission of three supernova remnants require colder dust temperatures than obtained by previous theoretical predictions. Their findings highlight the necessity of factoring in the diversity of inter- and circumstellar phases that dust grains evolve in.                       

At Solar System scales, dust grains serve as the building blocks for planetary systems \citep[see review by][]{Johansen_2017}. Anomalies in the nucleosynthetic isotope composition of primitive presolar meteorite grains with respect to solar values reveal their stellar origin, and the nucleosynthetic fingerprint of these isotopes imparts key constraints on planet formation theory \citep[e.g.][]{Burkhardt_2011, Ek_2020, Nie_2023, Onyett_2023}. The carriers of $s$-process and $r$-process nuclides predominantly originate respectively from low-mass asymptotic giant branch (AGB) stars \citep{Gallino_1990, Straniero_2006} and explosive nucleosynthetic events such as neutron star mergers or supernovae \citep[][and references therein]{Cowan_2021}. These exhibit heterogeneity within meteorites with parent bodies sourced from the inner parts of the Solar System -- the noncarbonaceous (NC) group, depleted in $r$-process elements -- relative to outer Solar System bodies beyond Jupiter's orbit -- the carbonaceous (CC) group, enriched in $r$-process elements \citep[for a review on the NC-CC dichotomy see e.g.][]{Kruijer_2020, Bizzarro_2025}. 
The conventional view that the rapid growth of Jupiter led to a spatial separation that prohibited the mixing of material between the inner and outer reservoirs inside the protoplanetary disc has been challenged in recent work by \citet{Liu_2022} and \citet{Colmenares_2024}, who proposed instead that $r$-process isotopes are missing in inner Solar System bodies due to thermal processing of presolar grains by stellar outbursts. Our understanding of the early evolution of the Solar System is thus highly dependent on accurately tracing presolar carriers of $s$- and $r$-process back to their stellar origin, which is a strong motivation for studying the relative contributions from AGB stars and supernovae to star- and planet-forming dust. 

Inferred traces of the short-lived isotope $^{26}$Al (half-life $\tau_{1/2} \approx 0.72 \,\mathrm{Myr}$) are of particular interest in the study of presolar grains. Its radioactive decay provides an important heat source that contributes to the differentiation \citep{Hevey_2006} and dehydration \citep{Grimm_1993} of planetesimals, resulting in decreased radius and bulk water abundances of terrestrial planets \citep[see e.g.,][]{Lichtenberg_2019}. Excesses of $^{26}$Mg, the daughter isotope of $^{26}$Al decay, relative to solar abundances detected in presolar primitive meteorite grains and interplanetary dust particles also reflect the stellar nucleosynthetic environment of $^{26}$Al and provide valuable information about the mixing processes occurring in their parent stars \citep[e.g.][]{Nittler_1997, Zinner_2007, Hoppe_2023}. The canonical initial Solar System value of $^{26}$Al/$^{27}$Al $\approx$ $5 \times 10^{-5}$ is inferred for the oldest solids found within meteorites, the so-called calcium-aluminum rich inclusions (CAI) that condensed around the forming Sun 4,567 Myr ago \citep[][]{Connelly_2012}. High $^{26}$Al$/$$^{27}$Al ratios of up to $0.2$ were first found in presolar SiC grains in the Murchison meteorite \citep{Zinner_1991}, and have since also been inferred in presolar graphite \citep[up to $\sim$$0.1$,][]{Hoppe_1995, Jadhav_2013, Amari_2014} as well as oxide grains such as corundum \citep[up to $\sim$$10^{-3}$,][]{Hutcheon_1994, Choi_1998} and spinel \citep[up to $\sim$$0.01$,][]{Zinner_2005}. The comparable size of grains to the spatial resolution of ion probes such as NanoSIMS renders the research on presolar silicates, which cannot be chemically separated from the meteorites, particularly challenging and imprecise up until recently \citep[e.g.][]{Hoppe_2018}. O-rich presolar stardust had initially been inferred to condense primarily in the winds of red giant branch and AGB stars of nearly solar metallicity \citep[][]{Nittler_2009}. More recent results \citep{Hoppe_2021}, however, suggest that SNe may have contributed to $>$$30\%$ of presolar silicates. The supernova contribution to other minerals may be more important than initial estimates \citep[$>$$5\%$ for oxides, $\sim$$30\%$ for graphite, $\sim$$90\%$ for Si$_3$N$_4$, or in total $>$$25\%$ for all types; for a detailed review see][]{Hoppe_2022}. Nevertheless, the relative contribution of late-type giant stars and core--collapse supernovae to the solar nebula and hence the meteoritic dust inventory of the Solar System remains a matter of controversy \citep[see also][]{Dwek_1998, Zhukovska_2007}.          

The aim of this work is thus to explore the evolution of dust within the multi-phase nature of a Milky Way--like galactic setting and to compare this with the relative numbers of presolar grains from AGB stars and supernovae found in meteorites. The paper is organized as follows. In Section~\ref{sec:simulations} we outline the set-up and initial conditions of the galaxy simulation, the subgrid recipes used for star formation and stellar feedback, as well as the tracer particle implementation. We provide an overview of the methods used for the analysis, along with the dust evolution model in Section~\ref{sec:analysis}. Section~\ref{sec:results} is dedicated to evaluating the thermal sputtering efficiency for different categories of particles in the ISM, as well as the comparison between the evolution of AGB and SN dust grains and their link to presolar grains. The implications and limitations of our results are discussed in Section~\ref{sec:discussion} and summarized in Section~\ref{sec:conclusions}.

\section{Simulation}
\label{sec:simulations} 

\subsection{Simulation set-up}
\label{sec:sim}
The simulation analysed for the purposes of this paper was performed with the hydrodynamical and N-body code {\small RAMSES} \citep{RAMSES}, which solves the Euler equations in their conservative form with a second-order Godunov method employing a Riemann solver \citep[see][]{RAMSES}. It is evolved over the course of $t \sim 850 \, \mathrm{Myr}$. The equation of state is that of a mono-atomic ideal gas with an adiabatic index of $\gamma = 5/3$\footnote{We do not explicitly treat the formation of molecules such as H$_2$. More sophisticated models in which an effective equation of state is inferred exist \citep[see e.g.][]{Nickerson_2018}, but the large-scale dynamics of the galaxy would remain unaffected by a more advanced treatment.}. We track the cooling and heating of primordial gas using equilibrium thermochemistry \citep{Courty_2004}, accounting for collisional ionization and excitation, recombination, bremsstrahlung, Compton cooling and heating, and dielectronic recombination. The contribution of metal lines to the total cooling rate is extracted from tabulated models generated with CLOUDY \citep{CLOUDY}. We account for on-the-fly self-shielding \citep{Aubert_2010, Rosdahl_2012} and heating from a spatially uniform ultraviolet background (updated from \citet{Haardt_1996}; see \citet{Rey_2020}, for further details). Furthermore, the gas is heated by stellar feedback processes, described in Section \ref{sec:SF}.

Our set-up allows us to recover a turbulent, multi-phase ISM \citep[see e.g.][]{McKee_1977}, depicted in Fig.~\ref{fig:temp_dens_fg10}, with the existence of a hot phase ($\gtrsim$$10^6 \, \mathrm{K}$) being powered by supernova explosions. The warm ($\sim$$10^4 \, \mathrm{K}$) and cold ($<$$10^2 \, \mathrm{K}$) phases are regulated by cooling and heating processes, as well as gas compression and expansion. {\small RAMSES} is structured according to a Fully Threaded Tree and uses an Adaptive Mesh Refinement (AMR) scheme. We split cells into 8 new cells when they contain 8 dark matter particles, and when their baryonic mass (star particles and gas) exceeds $8\,m_{\rm bar}$, where $m_{\rm bar}= 1175\,{\rm M}_{\odot}$. The finest cell size reached is $\Delta x\approx 9 \, \mathrm{pc}$. At this spatial and mass resolution, we are able to resolve the locations and bulk properties of molecular clouds, but not their internal structure \citep[][]{Grisdale_2018,Grisdale_2019}. This is sufficient for our purposes, but could be improved upon in future work (see Section~\ref{sec:resolution}).

\subsection{Initial conditions}
\label{sec:ics}
The simulated galaxy represents a Milky Way progenitor without a bar at a redshift of $z \sim 0$, embedded in a dark matter (DM) halo of mass $M_{\text{h}} = 1.25 \times 10^{12} \, \text{M}_{\sun}$ following a Navarro--Frenk--White profile \citep{Navarro_1996}. The virial radius and mass of the DM halo are $r_{200} = 205 \, \mathrm{kpc}$ and $M_{200} = 1.1 \times 10^{12} \, \text{M}_{\sun}$, respectively. The galaxy has an initial gas fraction of $f_{\text{g}} = 10\%$ and metallicity $Z = 1.67 \, Z_{\sun}$, with the gas fraction defined as
\begin{equation}
    f_{\text{g}} = \frac{M_{\text{g}}}{M_{\star} + M_{\text{g}}},
\end{equation}
i.e. the ratio of mass in gas to the total stellar and gas mass. Stellar particles are included in the beginning of the simulation, such that our galaxy resembles the Milky Way, which has a gas fraction of around 10 per cent \citep[e.g.][]{Nakanishi_2016, McMillan_2017}. These initial particles represent very old stellar populations and are passive in the sense that they are treated as dark matter particles by the code and thus do not contribute to the feedback processes described in Section~\ref{sec:SF}.\footnote{Implementing an age distribution for these initial particles would yield a more realistic set-up of the galaxy, but treating them as passive with respect to feedback is sufficient for our purposes.}    
The initial conditions were based on the AGORA project \citep{AGORA_2014}. The bulge has a mass of $M_{\text{b}} = 4.3 \times 10^{9} \, \text{M}_{\sun}$, a scale radius of $r_{\text{b}} = 0.3432 \, \mathrm{kpc}$ and follows a \citet{Hernquist_1990} surface density profile. The stellar disc mass is $M_{\text{d}} = 3.87 \times 10^{10} \, \text{M}_{\sun}$ and follows an exponential profile given by:  
\begin{equation}
    \rho(r, z) = \rho_{\text{d}} \exp(-r/r_{\text{d}})\cdot \exp(-|z|/z_{\text{d}}), 
    \label{eq:rho_disc}
\end{equation}
where $r_{\text{d}}= 3.43 \, \mathrm{kpc}$ and $z_{\text{d}}= 0.34 \, \mathrm{kpc}$ are the scale radius and height of the disc, respectively, and $\rho_{\text{d}} = M_{\text{d}}/4\pi r_{\text{d}}^2 z_{\text{d}}$. The galaxy initially contains approximately $4.8 \times 10^5$ star particles, and about four times as many star particles after running the simulation for $\sim$$850 \, \mathrm{Myr}$. The initial mass in gas is approximately $4.3 \times 10^9 \, \mathrm{M_{\sun}}$. With a star-formation rate of $\sim$$1 \, \mathrm{M_{\sun} \, yr^{-1}}$ maintained during $1 \, \mathrm{Gyr}$, about $10^9 \, \mathrm{M_{\sun}}$ of gas are converted into stars during that time period. After $1 \, \mathrm{Gyr}$, the gas fraction $f_{\text{g}} \approx 8\%$, meaning the resemblance of the galaxy to the Milky Way is closer at early times but still reasonably close throughout the $\sim$$850 \, \mathrm{Myr}$ during which we evolve it. 

The gas is initialized on the AMR grid and is evolved by solving equations (8) -- (10) in \citet{RAMSES}. Initially, the disc is isothermal and entirely supported by rotational and thermal energy. Without any pre-existing regulation by feedback or gravity-driven turbulence, introducing the cooling and star formation processes described in Section~\ref{sec:sim} and \ref{sec:SF} from the beginning of the simulation will result in unphysical fragmentation of the disc, as the cooling time of the gas is much shorter than the dynamical time-scale. We therefore evolve the simulation without cooling or star formation (and thus no feedback) for $t_0\approx145 \, \mathrm{Myr}$, corresponding to approximately one orbital period at a galactocentric radius of $5$ kpc. This `relaxation period', which was chosen based on when large scale spiral structure had visually formed, allows for gravity-driven turbulence to develop, which softens the transition into a galaxy that is supported by turbulence and thermal energy. For all subsequent analysis, we neglect the data produced prior to $t_0\approx145 \, \mathrm{Myr}$, and consider $t=0$ as the first time step in the simulation after this time period. At the end of the relaxation period, the gas is nearly in hydrostatic equilibrium -- with some turbulent motions stabilising the disc. However, the equilibrium is lost when the cooling and feedback processes detailed in Section~\ref{sec:sim} and \ref{sec:SF} are initiated. A new quasi-equilibrium, where the gas can collapse and expand locally, is achieved after slightly more than one rotational period ($\sim$$0.1 \, \mathrm{Gyr}$). At this stage, the global rate of star formation has reached $\sim 2~{\rm M_\odot/yr}$, which is maintained throughout the simulation runtime. The ISM pressure is supported by turbulence and thermal energy.\footnote{We do not consider magnetohydrodynamics. As such, magnetic pressure does not contribute to the pressure support of the ISM in our simulation (see Section~\ref{sec:discussion}).}

\subsection{Star formation and stellar feedback}
\label{sec:SF}
The subgrid recipes for star formation and stellar feedback follow the same prescription as in \cite{Agertz_2021} \citep[see also][]{Agertz_2013}. Star formation occurs as a Poisson random process, sampled at each simulation time step by a star formation rate density of 

\begin{equation}
    \dot{\rho}_{\star} = \frac{\rho_{\text{g}}}{t_{\text{SF}}}, \quad  \, \text{if} \, \rho_{\text{g}} > 100 \, \mathrm{amu\, cm^{-3}}.
    \label{eq:SFR}
\end{equation}
In equation \ref{eq:SFR}, $\rho_{\text{g}}$ is the cell gas density and $t_{\text{SF}} = t_{\text{ff}} / \varepsilon_{\text{ff}}$ with $t_{\text{ff}} = \sqrt{3\pi/32G\rho_{\text{g}}}$ the local free-fall time and the star formation efficiency per free-fall time $\varepsilon_{\text{ff}} = 10\%$. This choice of $\varepsilon_{\text{ff}}$, combined with our feedback scheme described below, has been shown to reproduce global galaxy properties \citep[][]{Agertz_2015}, as well as giant molecular cloud (GMC) characteristics -- such as size and mass distributions, and star formation efficiencies -- that closely match observations of the Milky Way \citep[][]{Grisdale_2018, Grisdale_2019}.
An additional temperature criterion is adopted with a threshold at $T = 100 \, \mathrm{K}$ such that star formation occurs in the cold and dense ISM, in alignment with observations \citep[e.g.][]{Bigiel_2008}. 
When equation~\ref{eq:SFR} is fulfilled in a given cell, $N$ star particles are created according to a Poisson random process \citep[][]{Agertz_2013}. This stochastic sampling is needed to reproduce the star formation rate of a Milky Way-like galaxy, as sampling equation~\ref{eq:SFR} at every fine time step in the simulation would result in an overproduction of star particles. Stellar particles in the simulation are formed with initial individual masses of $m_{\star} = 10^3 \, \text{M}_{\sun}$, or multiples thereof ($N m_{\star}$) if $N>1$. Each star particle represents a stellar population with an associated \citet{Chabrier_2003} IMF. 

Stellar feedback processes include Type II and Type Ia supernovae (SNII and SNIa), radiation pressure, and stellar winds from both asymptotic giant branch stars and massive stars ($M \gtrsim 8 \, \mathrm{M_{\sun}}$). These processes inject energy, momentum, mass, and metals -- at rates dependent on stellar evolution -- into the nearest grid point \citep[for details, see][]{Agertz_2013}. The feedback depends on stellar age, mass, and gas/stellar metallicity through the metallicity-dependent age–mass relation of \citet[][]{Raiteri_1996}, calibrated using the stellar evolution code {\small STARBURST99} \citep[][]{Leitherer_1999}. Only the star particles that are formed when the condition in equation~\ref{eq:SFR} is fulfilled take part in these feedback mechanisms, i.e. not the primordial disc particles from the initial conditions described in Section~\ref{sec:ics}.

We track iron (Fe) and oxygen (O) abundances separately, as scalar fields, using yields from \citet[][]{Woosley_2007}. To compute the gas cooling rate, which depends on the total metallicity, we construct a total metal mass in each cell as $M_{\rm Z} = 2.09\,M_{\rm O} + 1.06\,M_{\rm Fe}$, following the solar abundance pattern of $\alpha$-elements (C, N, O, Ne, Mg, Si, S) and iron-group elements (Fe, Ni) from \citet[][]{Asplund_2009}. Metals are advected as passive scalars, i.e. we assume that the motion of metals in the gas and in the dust are well coupled. As the mean free path of the gas is much longer than the dust grain size on the spatial scales resolved in our simulations ($\gtrsim$$10$ pc), the length scale over which dust and gas decouple is determined by the stopping time of a grain in the Epstein regime \citep[][]{Epstein_1924} and the velocity dispersion of the gas in the ISM\footnote{The stopping time is given by $t_{\text{s}} \sim \rho_{\text{grain}} a/(\rho_{\text{g}}c_{\text{s}})$, where $\rho_{\text{grain}} \sim 1 \, \mathrm{g \, cm^{-3}}$ is the grain density and $c_{\text{s}}$ is the gas sound speed \citep[e.g.][]{Commercon_2023}. The spatial coupling scale of dust and gas is then given by $l_{\text{s}} \sim t_{\text{s}} \cdot v_{\text{d}}$, where $v_{\text{d}}$ is the relative velocity between dust and gas, which can be estimated through the velocity dispersion of the gas in the ISM.}. The velocity dispersion of the gas in the diffuse atomic ISM (HI) is on the order of $\sim$$10 \, \mathrm{km \, s^{-1}}$ on $\mathrm{kpc}$ scales, and on the order of several $\mathrm{km \, s^{-1}}$ on $\sim$$100 \, \mathrm{pc}$ scales \citep[see e.g.][]{Roy_2008, Agertz_2015_b}. Typical densities of $\rho_{\text{g}} \sim 10^{-21}$--$10^{-25} \, \mathrm{g \, cm^{-3}}$ between cold molecular gas and diffuse atomic HI in the ISM \citep[][]{McKee_1977}, along with the associated velocity fluctuations, imply that dust and gas can only decouple over scales far below $10 \, \mathrm{pc}$, and at most on scales of $\sim$$10 \, \mathrm{pc}$ in the diffuse ISM.\footnote{With these density and velocity dispersion values we find a spatial coupling scale of $l_{\text{s}} \sim 0.1 \, \mathrm{pc}$ for the cold molecular ISM and $l_{\text{s}} \sim 10 \, \mathrm{pc}$ for the diffuse atomic ISM, assuming a grain size of $a \sim 0.1 \, \mathrm{\micron}$ and  $c_{\text{s}} \sim 0.1$ -- $1 \, \mathrm{km \, s^{-1}}$.} Since we resolve turbulent motions only down to several 10s of $\mathrm{pc}$, the assumption of dust-gas coupling is a reasonable approximation.

Supernovae are assumed to inject $10^{51} \, \mathrm{erg}$ of thermal energy in discrete events. When the resolution is sufficient to resolve the SN cooling radius with at least three grid cells, the evolution of the remnant can be modelled self-consistently. In this regime, the hot bubble performs work on its surroundings, generating momentum in the gas. However, if the resolution is insufficient, the injected thermal energy is rapidly radiated away, leading to an underestimation of the impact of SN feedback. To mitigate this, we initialize such SNe in the momentum-conserving phase, injecting the full post–Sedov--Taylor momentum into the ISM \citep[see][]{Kim_2015}.

\begin{figure*}
    \includegraphics[width=2\columnwidth]{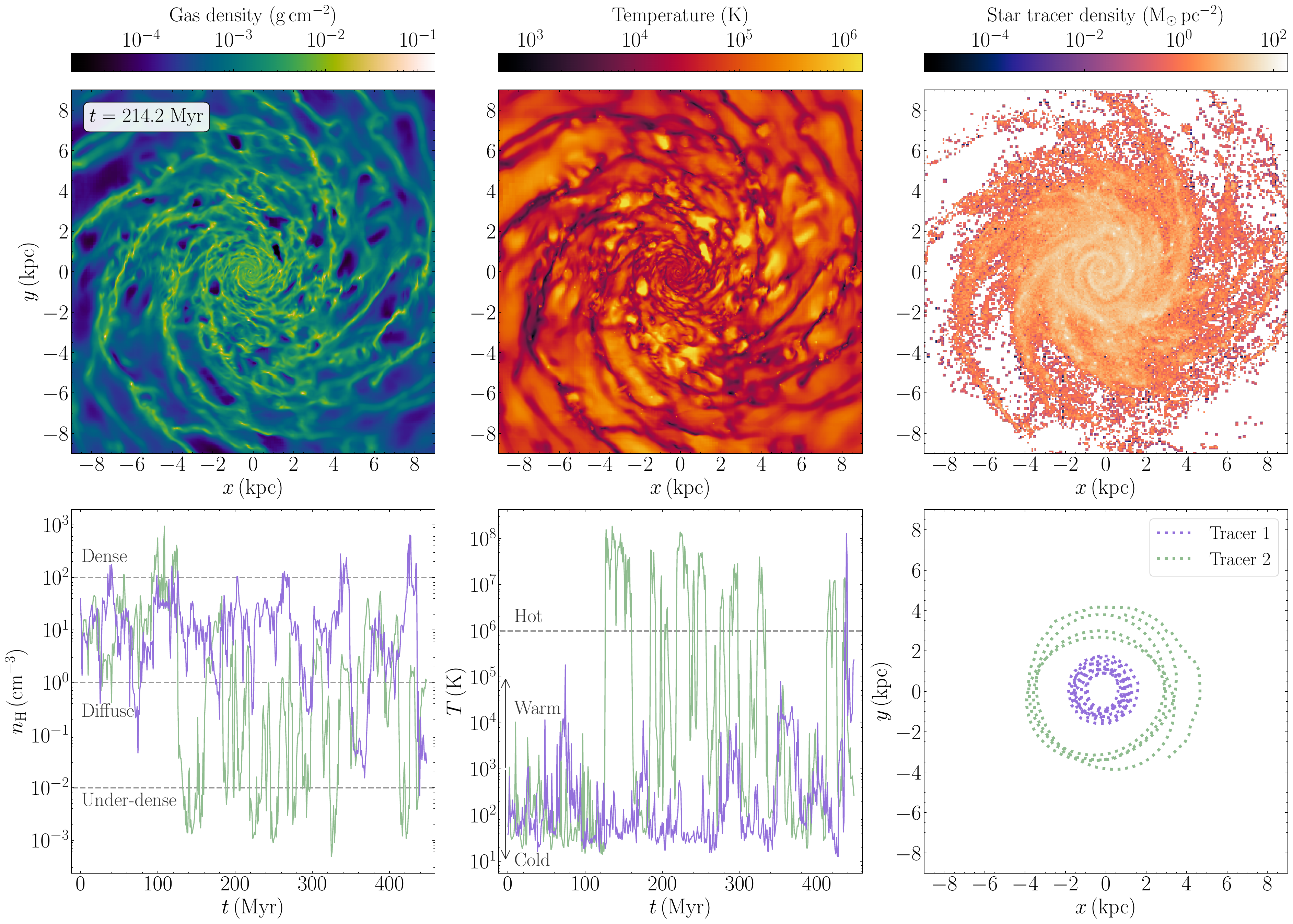}
    \caption{Top-row: Projected gas density (left), temperature (centre), and star tracer density (right) of the simulated galaxy seen face-on at a time of $214.2 \, \mathrm{Myr}$. Hot, under-dense pockets are carved out by stellar feedback and SN explosions. The tracers attached to star particles sample the population of star particles in the simulation. Not all star particles host tracers, but the top right panel shows tracers which are in a star state in this simulation snapshot and may be released (i.e. transition to gas tracers) later on in the simulation through feedback processes. Bottom-row: number density (left), temperature (centre) and position (right) evolution for two tracer particles as a function of time $t$ in the simulation during $450 \, \mathrm{Myr}$ of galactic evolution. These illustrate that even though the particles travel on nearly circular orbits around the galaxy, they are exposed to abrupt changes in phases and traverse both dense and cold regions of the ISM, as well as the hot and under-dense ionized medium. The various ISM phases are shown as density and temperature intervals for reference. The cold neutral medium is associated with typical densities of $n_{\text{H}} \gtrsim 10 \, \mathrm{cm^{-3}}$, but our label "Dense" refers to star-forming regions (see Section~\ref{sec:SF}).}
    \label{fig:temp_dens_fg10}
\end{figure*} 

\subsection{Tracing the evolution of baryonic matter}
\label{sec:tracers}
The advection of passive field scalars across grid cells, such as the O and Fe fields described above, leads to mixing of the gas, which prevents the retracing of the origin of individual gas parcels \citep[for a review on the numerical implementation of Eulerian schemes that include passive scalars in galaxy simulations, see][]{Teyssier_2015}. Moreover, phase changes between gas and stars in the processes of star formation and stellar feedback cannot be followed with passive scalar fields. In order to sample the motion and track the evolution of the baryons throughout the duration of the simulation, we therefore use a version of {\small RAMSES} that includes Lagrangian tracer particles, based on the Monte Carlo technique implemented by \citet{Cadiou_2019}. This implementation accurately reproduces the gas evolution in classical tests for the motion of the gas such as the Sedov--Taylor explosion or the Kelvin--Helmholtz instability \citep[see][for details]{Cadiou_2019}. The tracers each represent a fixed mass of $m_{\text{t}} \approx 4654 \, \mathrm{M_{\sun}}$, such that they sample the baryonic matter distribution but do not contribute to additional gravity and can thus be considered massless. A total of $2.42 \times 10^6$ tracer particles are present from the beginning of the simulation. After evolving the simulation for $\sim$$850 \, \mathrm{Myr}$, $2.86 \times 10^5$ tracers are attached to a star particle.   

Tracer particles that are not attached to a star particle sample the motion of the gas inside the galaxy and therefore move on nearly circular orbits. None the less, they undergo rapid transitions between ISM phases, moving from (to) cells containing hot, low-density plasma ($T \gtrsim 10^6 \, \mathrm{K}$; $n_{\text{H}} \lesssim 10^{-2} \, \mathrm{cm^{-3}}$) to (from) cold and dense gas ($T \lesssim 10^2 \, \mathrm{K}$; $n_{\text{H}} \gtrsim 10^{2} \, \mathrm{cm^{-3}}$) on time-scales of $\sim$1--$10 \, \mathrm{Myr}$. An example for the evolution of two tracers during a time interval of $450 \, \mathrm{Myr}$ in the simulation is shown in the bottom row of Fig.~\ref{fig:temp_dens_fg10}.     

During the processes of star formation and stellar feedback detailed in Section~\ref{sec:SF}, the tracer particles may transition between gas and star states. If a gas tracer is located in a cell where the criterion in equation~\ref{eq:SFR} is fulfilled, it can become attached to a newly formed star particle. Inversely, a tracer attached to a star particle undergoing mass-loss or exploding as a supernova can be converted into a gas tracer. The implementation for the transfer of tracers between gas and star states -- both star formation and mass-loss -- is probabilistic. If a fraction of gas $ p = m_{\star}/m_{\text{g}}$ is depleted in the process of star formation in a cell containing a gas mass of $m_{\text{g}}$ (sampled by tracers), then each tracer has a probability $p$ of becoming attached to the new star particle \citep[for further details, see section 2.2 in][]{Cadiou_2019}. The Monte Carlo tracer scheme converges to the underlying spatial distribution of stellar and gas mass, as well as mass fluxes between the two in the continuum limit. This approximation is reasonable in our case due to the $\sim$$10^6$ tracers in the simulation, as well as the $\sim$$10^5$ star formation and mass-loss events taking place. We treat gas tracers released from stars as dust grains and follow their evolution from their ejection from a star to their reincorporation into a newly formed star, determining the grain properties at each output under the assumption that the dust is well-coupled to the gas. Hence the process of a particle transitioning from gas to star is termed `astration'.

A database for analysing the evolution of the tracer particles from the {\small RAMSES} outputs was set up using the {\small YT} package \citep{YT_2011}. The physical properties of the tracers -- gas or star -- are assigned according to the values inside their host cells. Tracers are identifiable by unique IDs, which remain constant throughout the simulation time -- even as they transition between gas and star states. The evolution of each particle can thus be tracked. The database contains information about the location and velocities of the massless tracer particles, as well as the gas density, temperature, mass and pressure inside their host cells at intervals of 0.3 -- $1.1 \, \mathrm{Myr}$ over the course of the simulation (approximately $850 \, \mathrm{Myr}$). On top of these data, we apply the dust evolution models described in Section~\ref{sec:analysis}.

In our adopted feedback model, core--collapse supernovae inject energy and momentum into the ISM during $\sim$4.6 -- $40 \, \mathrm{Myr}$ after star formation (see Section~\ref{sec:SF}). The upper limit of $40 \, \mathrm{Myr}$ corresponds to the main sequence lifetime of $\approx$$8 \, \text{M}_{\sun}$ stars. Below this mass, stars are no longer eligible to undergo core--collapse, but are assumed to enter the AGB phase after the main sequence. We therefore assume tracer particles released at a time $t <40 \, \mathrm{Myr}$ after the birth of their stellar cluster to be SN dust particles, whilst those released after $t \geq 40 \, \mathrm{Myr}$ are labelled as AGB dust particles. We note that particles released due to mass-loss in binary stars (SNIa) cannot be filtered out using our time-scale argument. However, with the adopted feedback prescription only around $15\%$ of supernovae are expected to be of type Ia over the course of $10 \, \mathrm{Gyr}$ \citep{Agertz_2013}, meaning their contribution is negligible relative to that of core--collapse supernovae and AGB stars in the $\sim$$0.8 \, \mathrm{Gyr}$ evolution of our simulation. Fast stellar winds from massive stars of masses $\gtrsim$$8 \, \text{M}_{\sun}$ also take place during $t < 40 \, \mathrm{Myr}$. What we label as SN dust is thus a combination of particles released in supernova explosions and in the rapid winds of their progenitors. All stars with $M \gtrsim 8 \, \text{M}_{\sun}$ will explode as supernovae over the course of $40 \, \mathrm{Myr}$ and this distinction is therefore of minor importance.

\section{Analysis and Methods}
\label{sec:analysis}          
Using the database containing the information about the tracer particles outlined in Section~\ref{sec:tracers}, the data are filtered such as to consider only the tracers that have been in the star state at some point throughout the simulation. We follow the evolution of the particles after having been released from a star, by applying the models detailed in the following subsections.

\subsection{Grain size distribution}
\label{sec:grain-size}
At release, the grains are assigned an initial grain size drawn from a power-law distribution. The fit to the interstellar extinction curve by \cite{Mathis_1977} suggests a power-law size distribution ${\text{d}}n/{\text{d}}a$ with an index of approximately $-3.5$. We weight that distribution by mass such that 
\begin{equation}
    m \frac{{\text{d}}n}{{\text{d}}a} \propto a^{-0.5},
\end{equation}
where $a$ is the grain size. \citet{Yasuda_2012} estimated that carbon-rich AGB stars will produce grains with sizes generally in the interval of 0.01--$0.1\, \micron$, whereas dust particles condensing in the ejecta of SNe type II may span several orders of magnitude in size but are not expected to exceed a few $\micron$ \citep[see e.g.,][]{Todini_2001, Nozawa_2003, Schneider_2004, Gall_2014}. The MRN distribution \citep{Mathis_1977} includes grains with sizes up to $0.25 \, \micron$. We therefore set $a$ to range between $a_{\text{min}} = 0.01 \, \micron$ and $a_{\text{max}} = 0.25 \, \micron$. We discuss the impact of these assumptions further in Section~\ref{sec:discussion-grain-size}.

\subsection{Dust destruction}
\label{sec:sputtering}
Dust grains are primarily destroyed or `atomised' in supernovae and shocks in the interstellar medium, through sputtering, i.e. grain erosion due to collisions with highly energetic atoms and ions. We adopt the sputtering rate approximation from \citet[][their equation (25.14)]{Draine_2011},   
\begin{equation}
    \frac{{\text{d}}a}{{\text{d}}t} \approx -10^{-6} \mathrm{\mu m \, yr^{-1}} \bigg[1 + \left(\frac{10^6 \, \mathrm{K}}{T}\right)^3\bigg]^{-1}\left(\frac{n_{\text{H}}}{\mathrm{cm^{-3}}}\right) ,
    \label{eq:sputtering_eq}
\end{equation}
for grains in a hot plasma of temperature $T$ and number density $n_\text{H}$. Equation~\ref{eq:sputtering_eq} is numerically integrated throughout the simulation time. We only consider the thermal sputtering of particles after they are released from a star until they become attached to a star particle again (if they do so). This means that the grain size $a$ is reset when a particle undergoes astration. Following \citet{Esmerian_2022}, we assume that thermal sputtering is a dominant dust destruction channel in the ISM, but note that we neglect other physical processes that may influence the evolution of dust, such as kinetic sputtering \citep[see e.g.][]{Hu_2019}. In view of our assumption of dust--gas coupling, as well as the sparse time stepping of our simulation, we do not capture the short time-scales on which dust grains encounter the SN shock and are exposed to non-thermal sputtering, although thermal sputtering remains active in shock-heated regions long after the shock itself has dissipated. An alternative implementation used by \citet{Dubois_2024} is to assume a grain size-dependent destruction efficiency by SNe \citep[see e.g.][]{Hirashita_2019}. We refrain from following their prescription due to the resulting parameter degeneracies, as well as the wide disagreement in the literature on the efficiency of dust destruction by supernovae described in Section~\ref{sec:intro}. It should be noted, however, that we thus cannot quantify precisely the extent to which supernova grains are atomised prior to their injection into the ISM. 

Direct sublimation should not play a significant role, as even dust grains in a high-temperature plasma of $T\sim10^6$--$10^8 \, \mathrm{K}$ are expected to remain at temperatures considerably lower than their sublimation temperatures, since dust grains effectively cool through infrared emission of the radiation they absorb \citep[see][]{Dwek_1987, Kobayashi_2011}. Grain growth mechanisms are also not taken into account here, as we are interested in the disruptive effects of dust grain processing in the galaxy. However, we discuss some of the ways in which the inclusion of dust growth might affect our results in Section~\ref{sec:growth}. Since we do not resolve structures below $10 \, \mathrm{pc}$, we cannot factor in the shielding of grains inside cold, dense clumps, nor their destruction in small pockets of hot gas. The sub-grid modelling of such effects could be introduced in post-processing, by e.g. assigning a lognormal density distribution inside each cell, as expected from turbulence in the ISM \citep[e.g.][and references therein]{Berkhuijsen_2008}. This would be interesting to investigate in future work, but goes beyond the scope of this paper.

\subsection{Supernova remnant models}
\label{sec:SNRmodels}
In the Sedov--Taylor stage following a supernova explosion, the forward shock evolves adiabatically until it has decelerated enough for the cooling time of the shocked gas to become shorter than the expansion time-scale of the shock wave \citep[e.g.][]{Ostriker_1988}. The supernova remnant transitions to a pressure-driven phase, in which radiative losses become important, forming a thin dense shell of rapidly cooling gas behind the shock. The time-scale for the adiabatic-radiative transition to occur was calculated by \citet{Blondin_1998}:
\begin{equation}
    t_{\text{tr}} \approx 2.9 \times 10^{4} E^{4/17}_{51} n^{-9/17}_0 \, \mathrm{yr}\, ,
    \label{eq:ttr_SNR}
\end{equation}
where $E_{51}$ is the energy of the explosion normalized to $10^{51}$ ergs and $n_0$ the upstream hydrogen number density in CGS units. Massive stars carve out their circumstellar medium by expelling the surrounding material through rapid winds, before undergoing core--collapse \citep[see e.g.][]{Tenorio-Tagle_1990,RogersPittard2013}. Supernovae are therefore expected to explode in highly under-dense regions of the galaxy, where $n_{\text{H}} \ll 1 \, \mathrm{cm^{-3}}$ \citep[see e.g.][]{Vovk_2015, Andersson_2023}. 

Due to our finest spatial resolution of $\sim$$9 \, \mathrm{pc}$, and minimal time interval of $\sim$$0.3 \, \mathrm{Myr}$ for generating outputs, we do not always resolve the details of the supernova remnant evolution. When a supernova explosion occurs in a dense region of the simulation, the SN bubble structure is below our resolution scales. In that case, the gas surrounding the supernova is not thermalised, and the cell temperature is therefore an underestimate of the gas temperature, although we ensure that the correct amount of momentum is injected into the ISM; see Section~\ref{sec:SF}. Along with these limitations, the considerable theoretical uncertainties on the efficiency of dust destruction in the supernova remnant motivate an exploration of varying models for its evolution. In addition to the fiducial model, in which we apply the dust evolution model detailed in Section~\ref{sec:grain-size} and~\ref{sec:sputtering} to the raw simulation data, we therefore introduce supplementary steps in post-processing for SN particles. In one case, we assume that all the produced dust lies in a cold and dense protected region at $T=10\, \mathrm{K}$ and $n_{\text{H}} \geq 1 \, \mathrm{cm^{-3}}$ (labelled CM), and in the other that the dust is subject to extreme heating to $T \geq 10^6 \, \mathrm{K}$ for varying densities and time-scales (labelled HM). The parameters for these model variations are provided and the models are evaluated in Section~\ref{sec:SNRvsISM}.

\section{Results}
\label{sec:results}

\subsection{Evolution in a multi-phase ISM}
\begin{figure*}
    \centering
    \includegraphics[width=2\columnwidth]{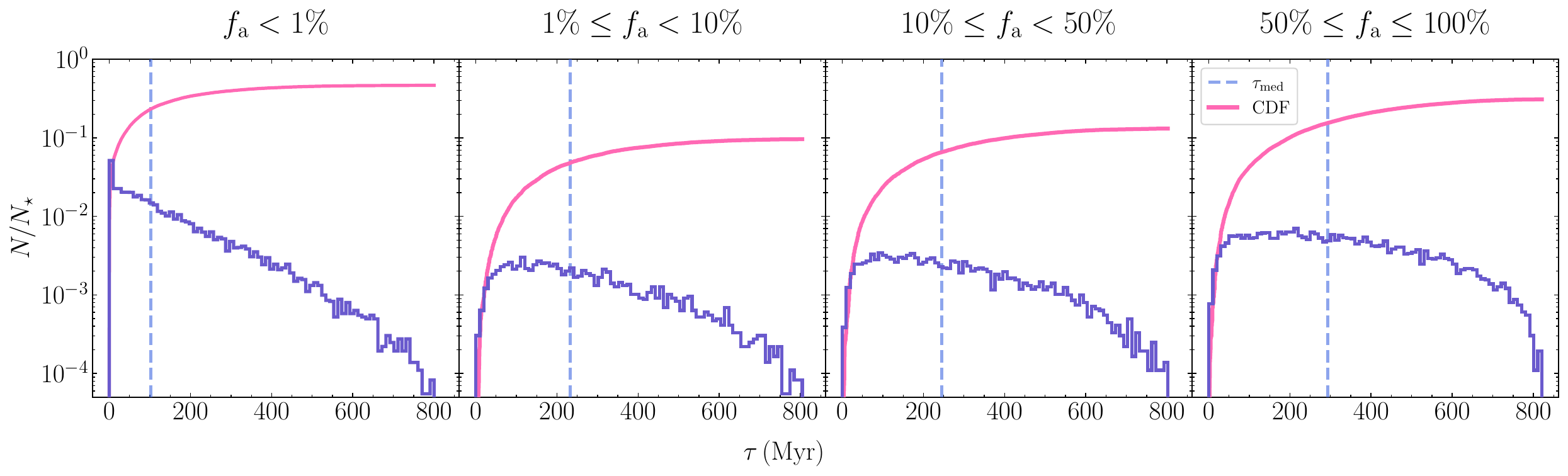}
    \caption{Amount of time $\tau$ between release from a star and new astration for dust particles in each category of atomisation fraction $f_{\text{a}}$. The histograms are binned in intervals of $10 \, \mathrm{Myr}$ and are normalized to the total number of astrating particles $N_{\star}$ (including all atomisation fraction categories). The solid lines represent the cumulative distributions and are also normalized to $N_{\star}$. The median free-floating time $\tau$ within each category is shown as a dashed blue line.}
    \label{fig:atomisation_cats_tau}
\end{figure*}
We begin by analysing the evolution of dust released from stars and the extent to which it is eroded by thermal sputtering on its interstellar journey. For this purpose, we separate the particles into four categories, depending on their final atomisation fraction. The atomisation fraction refers to the amount of grain mass lost due to thermal sputtering over the duration of the simulation. More specifically, we define the atomisation fraction $f_{\text{a}} = (a^3_{\text{0}} - a^3_{\text{f}})/a^3_0$ where $a_0$ and $a_{\text{f}}$ are the initial and final grain sizes, respectively. The final grain size $a_{\text{f}}$ of a particle is evaluated as its grain size at the end of the simulation, or at the time step before it is reincorporated into a star for particles that undergo astration. Note that we assume that atoms follow the same dynamics as the dust they originated from, such that even if dust is atomised the particles can still enter stars and planetary systems in their atomic form (and still represent their release from either supernovae or AGB stars).

To quantify the environmental impact of the galaxy on the survival of dust, we divide the grains into the atomisation fractions presented in Table~\ref{tab:fa_cats}. These categories include both the particles that reach the end of the simulation without undergoing astration, as well as the ones that are atomised prior to becoming attached to a star again. The time-scales between release and astration $\tau$ for star-forming particles are displayed in Fig.~\ref{fig:atomisation_cats_tau}. The large proportion of particles in categories (i) and (ii) in Table~\ref{tab:fa_cats} indicates that not only do almost half of all grains lose less than $10\%$ of their initial mass due to thermal sputtering in the interstellar medium, but a considerable fraction experiences, in fact, nearly no processing, with an atomisation fraction of less than $1\%$. The origin of this low degree of atomisation is twofold: grains are to a great extent released into cold regions of the ISM and are rapidly incorporated into star particles again -- on the order of $\leq$$10\, \mathrm{Myr}$ (see Fig.~\ref{fig:atomisation_cats_tau}). The values in Table~\ref{tab:fa_cats}, as well as the cumulative distributions shown in Fig.~\ref{fig:atomisation_cats_tau} indicate that nearly $50\%$ of all astrating grains are in category (i) -- i.e. they lose less than $1\%$ of their initial mass by thermal sputtering -- and that the majority of astrating particles in this category undergo astration less than $100 \, \mathrm{Myr}$ after their release from a star. Around $40\%$ of all tracer particles released from stars lose between $50\%$ and $100\%$ of their initial mass, but the CDF in Fig.~\ref{fig:atomisation_cats_tau} reveals that $<$$5\%$ of these become attached to a star particle again in the first $100 \, \mathrm{Myr}$ after their release. Their destruction can thus be attributed to their repeated exposure to the hot ISM, given the long free-floating time-scales, and many dust particles in category (iv) never undergo astration again. As a result, about $30\%$ of the astrating particles are in this atomisation fraction category, i.e. $\sim$$10\%$ less than the total proportion of particles that lose more than half of their initial mass due to thermal sputtering.
\begin{table}
	\centering
    \caption{Categories of particles based on their respective atomisation fractions, with the fraction of tracers in each category relative to the total number of tracer particles (middle) and to the number of astrating tracers (right).}
    \label{tab:fa_cats}
	\begin{tabular}{lcr}
		\hline
		     Atomisation fraction & All tracers & Astrating tracers \\
		\hline
		 (i) $f_{\text{a}} < 1\%$ & $34.8 \%$ & $46.5 \%$\\
        (ii) $1\% \leq f_{\text{a}} < 10\%$ & $10.7 \%$ & $ 9.6 \%$\\
        (iii) $10\% \leq f_{\text{a}} < 50\%$ & $15.4 \%$ & $ 13.1 \%$\\
        (iv) $50\% \leq f_{\text{a}} \leq 100\%$ & $39.1 \%$ & $ 30.8 \%$\\
        \hline
	\end{tabular}
\end{table}
\begin{figure*}
    \centering
    \includegraphics[width=2\columnwidth]{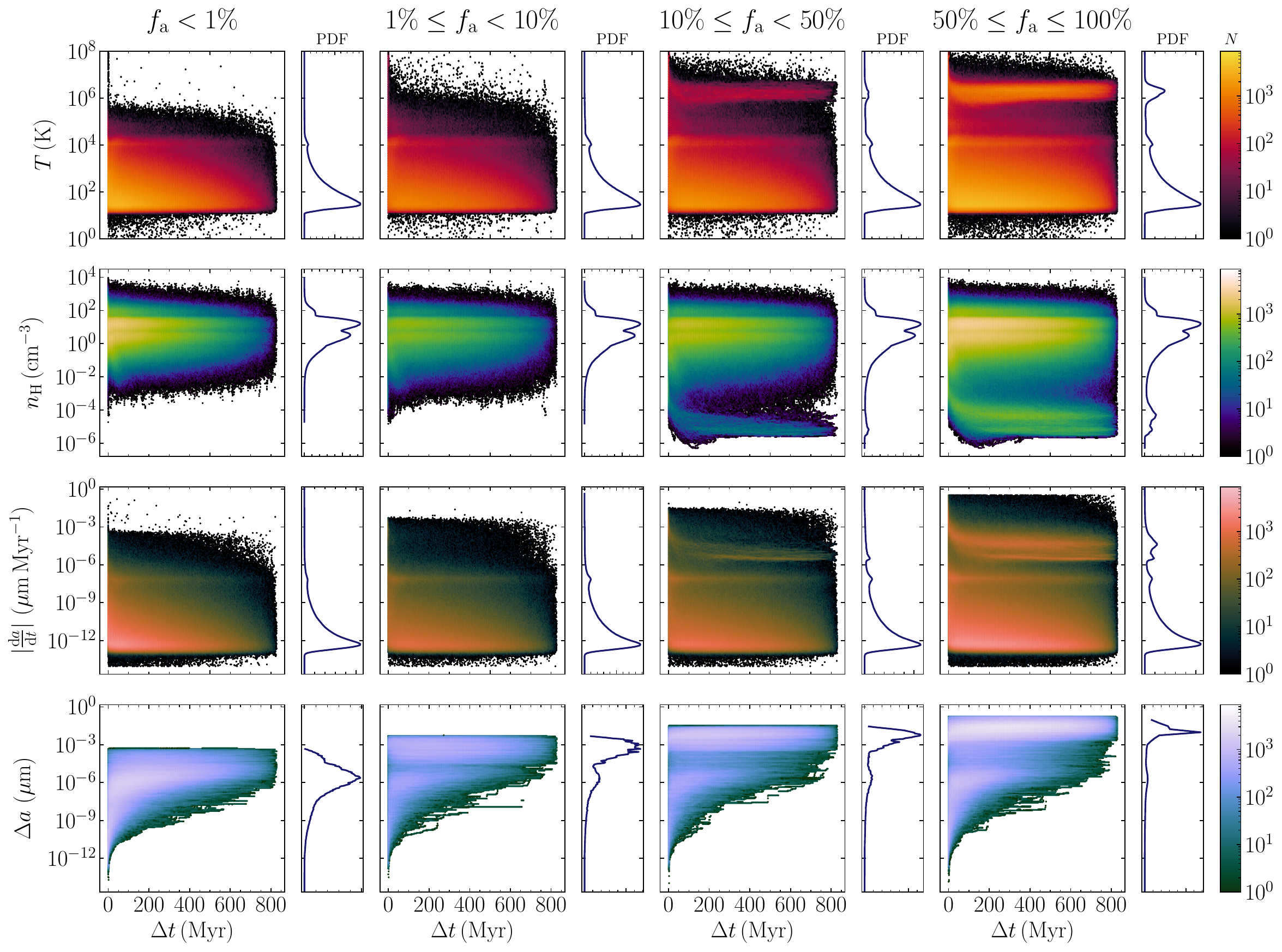}
    \caption{Dust evolution as a function of time after release from a star $\Delta t$, for dust grains losing less than $1\%$ ($1^{\text{st}}$ column), between $1\%$ and $10\%$ ($2^{\text{nd}}$ column), between $10\%$ and $50\%$ ($3^{\text{rd}}$ column) and between $50\%$ and $100\%$ ($4^{\text{th}}$ column) of their initial mass by the end of the simulation or prior to being incorporated into a star again. The first two rows depict the temperature $T$ and the number density $n_{\text{H}}$ of the ambient gas. The two bottom rows represent grain properties: the absolute value of the sputtering rate ${\text{d}}a/{\text{d}}t$, and the grain size reduction $\Delta a = a_0 - a(\Delta t)$ relative to the initial grain size $a_0$ after release. As we do not consider grain growth, all grains can only decrease in size and thus $|{\text{d}}a/{\text{d}}t| = - {\text{d}}a/{\text{d}}t$. The two-dimensional histograms are represented on a $250 \times 250$ grid, with $N$ the number of particles per bin. Marginal distributions are shown in the side panels to the right of each main panel, and are normalized such that the area under the curve is of unity.}
    \label{fig:atomisation_cats}
\end{figure*}

Fig.~\ref{fig:atomisation_cats} illustrates the evolution in time after being released from a star for the four families of dust grains. The two-dimensional histograms show the environments experienced by each category of grains in the simulation, as well as the resulting thermal sputtering rate and decrease in grain size, and highlight the multi-phase structure of the ISM in our simulation, with the gas spanning several orders of magnitude in number density and temperature. The survival or destruction of dust is shaped by its presence in the under-dense hot ionized medium (HIM; $T\gtrsim 10^6 \, \mathrm{K}$). Applying equation~\ref{eq:sputtering_eq} under these conditions yields a sputtering rate between $\text{d}a/\text{d}t = - 5 \times10^{-5} \, \mathrm{\micron \, Myr^{-1}}$ (for $T = 10^6 \, \mathrm{K}$; $n_{\text{H}} = 10^{-4} \, \mathrm{cm^{-3}}$) and $\text{d}a/\text{d}t = - 0.05 \, \mathrm{\micron \, Myr^{-1}}$ (for $T = 10^6 \, \mathrm{K}$; $n_{\text{H}} = 0.1 \, \mathrm{cm^{-3}}$). For increasing atomisation fractions, the particles progressively spend more time in the HIM, and the distribution of lifetimes for star-forming grains increasingly shifts towards longer time-scales (see Fig.~\ref{fig:atomisation_cats_tau}). The destruction of these particles is brought about by their exposure to highly heated regions of the galaxy, as well as the gradually longer duration of that exposure. Particles that end up being incorporated into a star again do so after predominantly several tens to hundreds of $\mathrm{Myr}$, which means they will have had time to cycle through a wide variety of phases (including the HIM) several times before being accreted by a star, as is exemplified in Fig.~\ref{fig:temp_dens_fg10}. The marginal distributions in Fig.~\ref{fig:atomisation_cats} none the less indicate that even the most severely sputtered grains in category (iv) spend most of their time in the cold neutral medium ($T\sim$ 10--$100 \, \mathrm{K}$; $n_\text{H} \sim$ 10--$100 \, \mathrm{cm^{-3}}$; $|\text{d}a/\text{d}t|\sim$$ 10^{-14}$--$ 10^{-11}\, \mathrm{\micron \, Myr^{-1}}$), suggesting that their complete destruction occurs during brief excursions to the HIM, on the order of $\sim$$1$--$10 \, \mathrm{Myr}$. Particles in category (iv) are also not predominantly subject to extreme temperatures at the time of release $\Delta t = 0$, which reveals that supernova grains do not represent the majority of the extremely eroded grains. These findings motivate a deeper investigation into the stellar sources responsible for the production of dust and how its survival is subsequently affected, which shall be explored in Section~\ref{sec:AGBvsSNe} and \ref{sec:SNRvsISM}.   

\subsection{AGB and SN dust sputtering in the interstellar medium}
\label{sec:AGBvsSNe}
\begin{figure}
    \centering
    \includegraphics[width=\columnwidth]{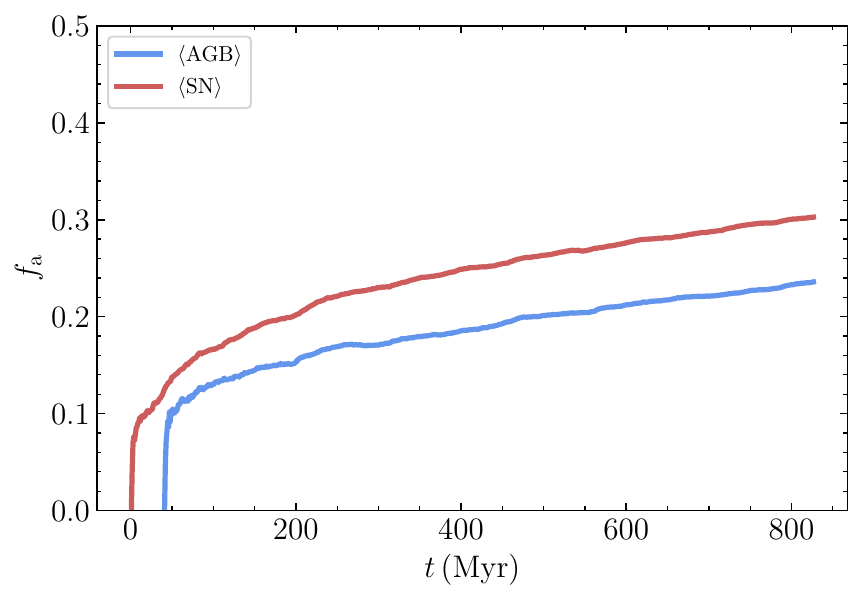}
    \caption{Evolution of the mean atomisation fractions of AGB (blue) and SN (red) particles over the duration of the simulation. The delay in the release of AGB dust particles, and thus also their atomisation, is due to the relatively slower evolution of low-mass stars. The mean values of $f_{\text{a}}$ remaining well below $50\%$ for both AGB and SN particles suggests that the grains preserve a significant proportion of their initial mass, on average, throughout the simulation.}
    \label{fig:meanfa_AGBvsSNe}
\end{figure}
As a next step, we distinguish particles that were produced in AGB stars from supernova grains, based on the criteria defined in Section~\ref{sec:tracers}. The mean values of the atomisation fractions for AGB and SN particles are plotted as a function of time $t$ in the simulation in Fig.~\ref{fig:meanfa_AGBvsSNe}. On average, supernova grains are more severely sputtered than AGB grains -- likely due to their release in shock-heated plasma following a supernova explosion. Yet their average degree of atomisation differs only by an additional $\sim$$10\%$ compared to AGB dust particles, and increases at most to $f_{\text{a}}\sim30\%$ after $800 \, \mathrm{Myr}$ of evolution in the simulation. The values are slowly increasing with time due to the presence of old dust particles that have been in circulation in the interstellar medium for several tens to hundreds of $\mathrm{Myr}$ without being reincorporated into a star.    
The mean atomisation fraction of AGB dust particles remains below $25\%$ at all times. This means that the majority of dust particles released from stars, both SN and AGB particles, cycle through the interstellar medium relatively unaltered by thermal sputtering, despite the presence of the hot ISM phase powered by supernova explosions -- in qualitative agreement with the results of \citet{Dubois_2024}. We next investigate whether this effect is the result of insufficient supernova bubble resolution in dense regions of the ISM, as described in Section~\ref{sec:SNRmodels}, and how varying the density and temperature environment of SN dust particles after release impacts their mass-loss due to thermal sputtering. As we are ultimately interested in linking our results to presolar grains, in the following subsections we only study the processing of particles that become attached to a star again after some time in the simulation.

\subsection{Destruction in the supernova remnant or in the ISM?}
\label{sec:SNRvsISM}
\begin{figure*}
    \centering
    \includegraphics[width=2\columnwidth]{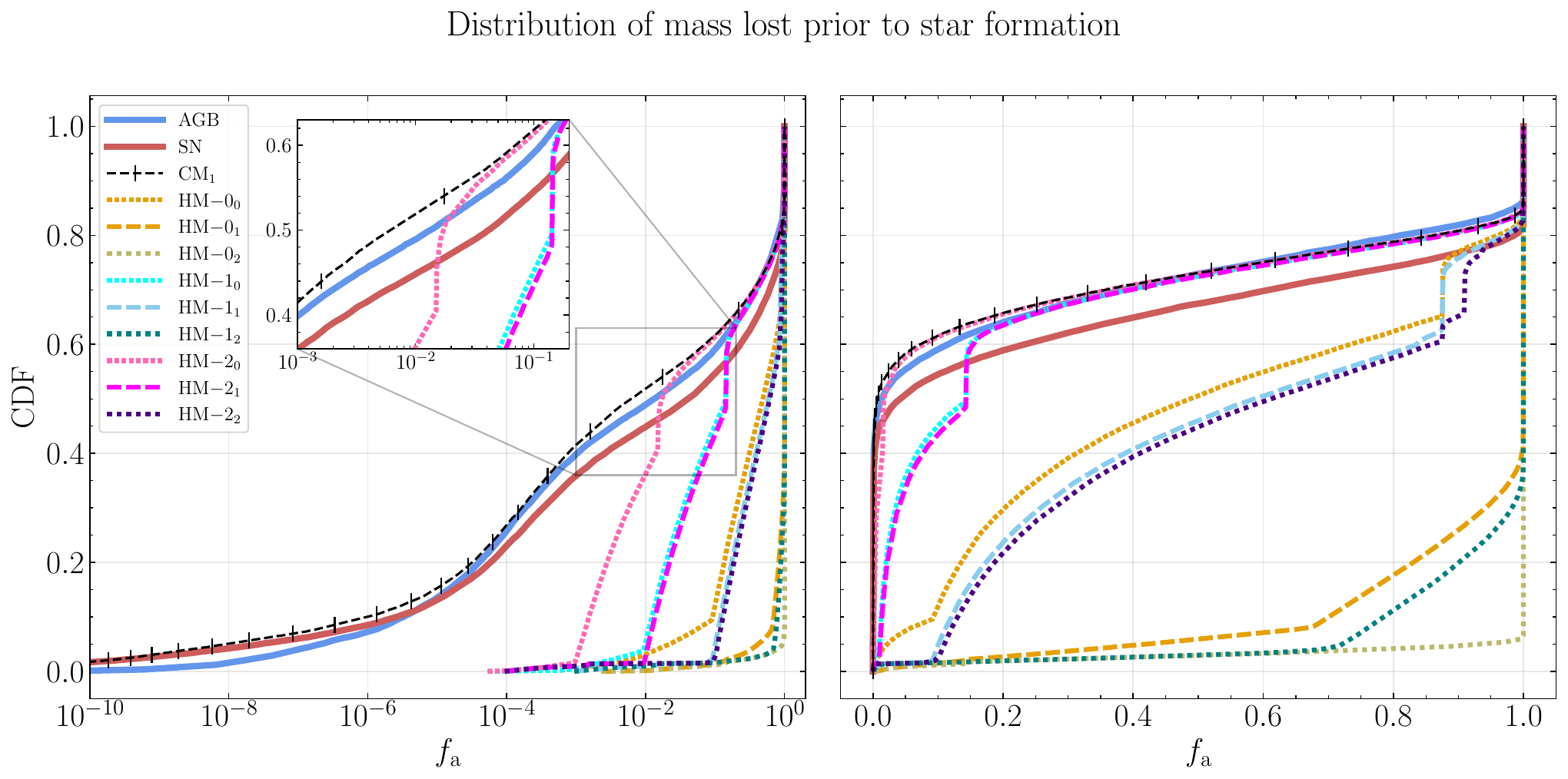}
    \caption{Cumulative distribution of mass lost by AGB and SN dust grains before undergoing astration. The left and right panels represent the same data but on logarithmic and linear scales, respectively. The fiducial models are represented by the solid lines, whereas the dashed/dotted lines display the model variations for the SNR evolution from Table~\ref{tab:models_table}. For readability, we do not to show the CM$_0$ and CM$_2$ lines as they are nearly identical and overlap very closely with with CM$_1$. The non-differentiable regions of the HM curves result from the transition in environment between the additional heating introduced in post-processing and the nominal temperature and density conditions encountered by SN particles in the simulation.} 
    \label{fig:models}
\end{figure*}
Examining the impact of their journey through the ISM reveals that the grains lose quite a small amount of their initial mass due to thermal sputtering before undergoing astration (Fig.~\ref{fig:meanfa_AGBvsSNe}), on average. We now go on to compare the evolution of AGB and SN dust particles that end up being reincorporated into a star. The cumulative distributions of the atomisation fractions for SN and AGB particles are shown in Fig.~\ref{fig:models}. In the nominal model, SN particles are subject to higher mass-loss fractions than AGB particles, but not exceedingly, as already established in Section~\ref{sec:AGBvsSNe}. About $10\%$ more AGB dust particles have an atomisation fraction of $\lesssim$$1\%$ compared to SN dust, and overall around $50\%$ of the supernova dust particles lose less than ten per cent of their initial mass due to thermal sputtering. 

We consider several model variations for the evolution of the dust inside the supernova remnant as well, as introduced in Section~\ref{sec:SNRmodels}, for which the parameters are shown in Table~\ref{tab:models_table}. The abbreviations CM and HM respectively stand for Cold Model and Hot Model. Our choice of parameters is motivated by equation~\ref{eq:ttr_SNR}, as well as the fact that supernovae are expected to explode in highly under-dense regions of the galaxy where $n_{\text{H}} \ll 1 \, \mathrm{cm^{-3}}$ (see Section~\ref{sec:SNRmodels}). In the scenario where $100\%$ of the dust lies in a dense and cold, shielded region of the swept-up supernova shell (CM$_{0, 1, 2}$), we obtain virtually indistinguishable atomisation fractions for all considered time-scales. In that case, the grain erosion history is very similar to that of AGB dust particles. On the other hand, if we assume the totality of newly condensed dust to be located in the shock-heated bubble regions, with $T \geq 10^6 \, \mathrm{K}$, the final degree of atomisation is highly dependent on the amount of time evaluated and on the density. The destruction of grains due to thermal sputtering is almost complete in 3 cases: HM$-0_{1, 2}$ and HM$-1_{2}$. The most extreme one, HM$-0_{2}$ with $100 \, \mathrm{kyr}$ spent in $1 \, \mathrm{cm^{-3}}$ density, we regard as highly unlikely, due to the transition to a pressure-driven snowplow phase on the order of $\sim$$10 \, \mathrm{kyr}$ for high densities of $\sim$$1 \, \mathrm{cm^{-3}}$ (see equation~\ref{eq:ttr_SNR}). More broadly speaking, the high density models HM$-0_{0, 1, 2}$ are probably not representative of true SNR physical conditions, since
\begin{enumerate}
    \item[(a)]SNe typically explode in very low density regions of the ISM, where $n_{\text{H}} \ll 1 \, \mathrm{cm^{-3}}$, as mentioned above. 
    \item[(b)]The sweeping up of matter by the SN shock further reduces the density.
    \item[(c)]The self-similar evolution of the upstream density for a Sedov--Taylor blast wave scales inversely with temperature \citep{Taylor_1950}. At temperatures $\geq$$10^6 \, \mathrm{K}$, densities of $1 \, \mathrm{cm^{-3}}$ are unphysically high.
\end{enumerate}
\begin{table}
	\centering
    \caption{Model parameters for the SNR evolution.}
    \label{tab:models_table}
	\begin{tabular}{lccr} % four columns, alignment for each
		\hline
		  Model & Time $(\mathrm{kyr})$ & $T \, (\mathrm{K})$ & $n_\text{H} \, (\mathrm{cm^{-3}})$\\
		\hline
		 CM$_{0,1,2}$ & 1, 10, 100 & 10 & $\geq 1$\\
		\hline
        HM$-0_{0,1,2}$ & 1, 10, 100 & $\geq 10^6$ & $\leq 1$\\
        HM$-1_{0,1,2}$ & 1, 10, 100 & $\geq 10^6$ & $\leq 0.1$\\
        HM$-2_{0,1,2}$ & 1, 10, 100 & $\geq 10^6$ & $\leq 0.01$\\
        \hline
	\end{tabular}
\end{table}
The physical parameters chosen for HM$-1_{2}$ are a more reasonable assumption, although the configuration in which all the dust is confined to the shock-heated plasma is still debatable \citep[see e.g.][]{Kirchschlager_2024a, Dedikov_2025}. In all other Hot Models, the sputtering efficiency is considerably reduced. For $\sim$$40\%$ of the SN particles in HM$-1_{0}$, HM$-2_{0}$ and HM$-2_{1}$, their atomisation fraction is on the order of a few per cent at most. Approximately $50\%$ of the grain population loses less than $50\%$ of the initial mass in all models except HM$-0_{1, 2}$ and HM$-1_{2}$. The proportion of grains that lose less than half of their initial mass is lower in these model variations than in the fiducial model by an absolute difference of $\sim$$20\%$.      

We emphasize that the physical modelling of core--collapse supernovae is multifaceted, and the inhomogeneous and dynamical structure of the supernova remnant implies that time-scale, density and temperature arguments are insufficient to make precise predictions about the outcome for dust grains condensing in its cooling ejecta. We cannot account for all disruptive physical processes to which the dust is exposed, and hence we possibly underestimate the extent to which it is atomised (see Section~\ref{sec:resolution}). However, here we aim to show that even in the most severe physical conditions, a considerable fraction of particles are not completely eroded away by thermal sputtering in the supernova bubble and even lose less than half of their original mass. Our estimate on the thermal sputtering efficiency in the nominal model might thus in reality be too conservative, but it is probably offset by an absolute difference of $\lesssim$$20\%$ at worst. None the less, we note that our model does not include the effects of non-thermal sputtering on dust (as outlined in Section~\ref{sec:sputtering}) and we cannot quantify precisely the degree of atomisation for SN particles. 

For the following part, we therefore consider the evolution of AGB and SN particles in relation to presolar grains, i.e. as they are reincorporated into a star, still assuming that they may undergo astration in their atomic form (independently of their degree of atomisation). This is also motivated by the fact that only $\sim$$4\%$ of astrating grains with ISM free-floating times of less than $100 \, \mathrm{Myr}$ lose more than half of their initial mass due to thermal sputtering (see Fig.~\ref{fig:atomisation_cats_tau}), and in total approximately $30\%$ of particles that undergo astration are in category (iv) in the fiducial model. The relative gas-phase depletion of refractory elements such as Si \citep[e.g.][]{Turner_1991} in dense molecular clouds implies that dust in its atomic form will likely become incorporated into the icy mantles that cover dust grains during the process of star formation \citep[see also][]{Ceccarelli_2018, Silsbee_2021}, meaning that atomised dust is still relevant for astration, as well as for planet formation.

\subsection{AGB and SN dust astration}
\label{sec:agb_SN_astration}

\begin{figure}
    \centering
    \includegraphics[width=\columnwidth]{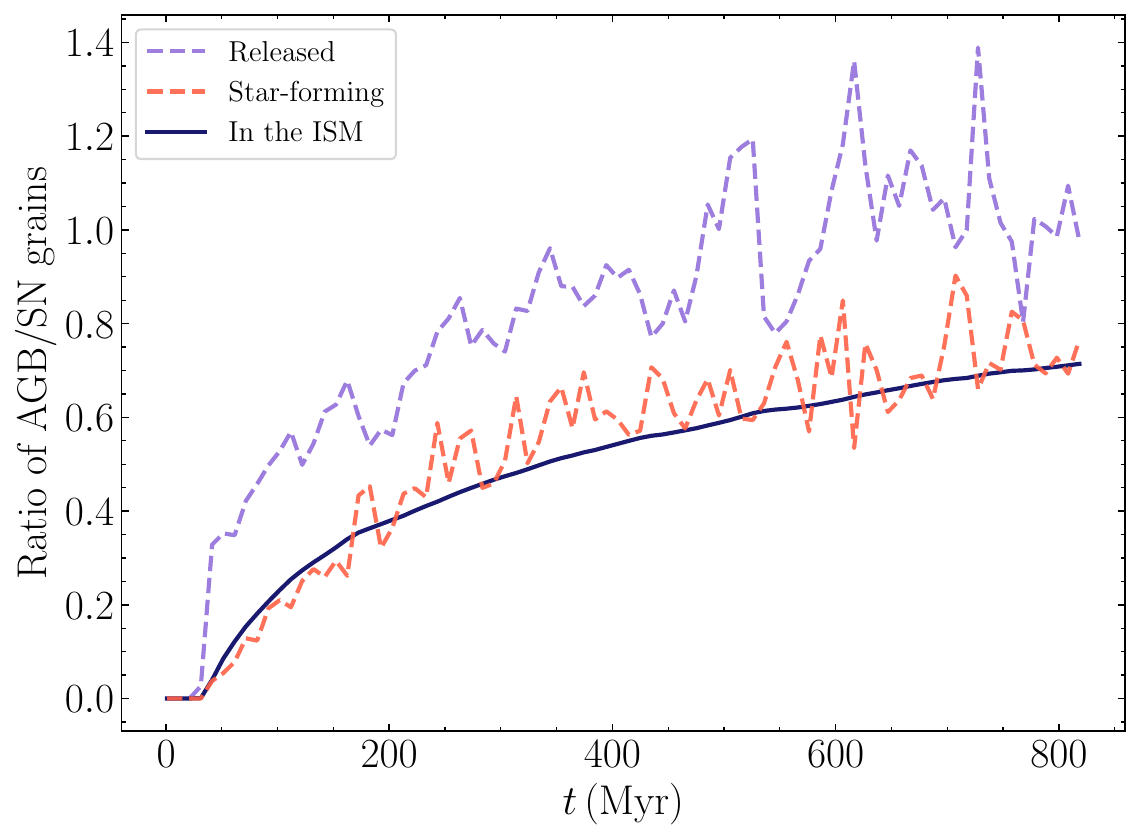}
    \caption{Relative amount of grains originating from AGB stars compared to supernovae as a function of time over the course of the simulation. The solid line represents the AGB/SN fraction of dust particles in circulation (i.e., not attached to a star particle) at any given time in the simulation. Dashed lines are respectively the proportion of released (purple) and astrating (orange) AGB/SN particles. We use time bins of $10 \, \mathrm{Myr}$.}
    \label{fig:AGBvsSNe}
\end{figure}

In order to assess the relative contribution of SN and AGB dust to newly formed star particles, we begin by comparing their corresponding release rates and overall abundances in the ISM throughout the duration of the simulation. The number ratios of AGB to SN dust particles released, astrating and in circulation in the ISM are shown as a function of time in the simulation in Fig.~\ref{fig:AGBvsSNe}. Early on in the simulation, mass-loss by supernovae dominates over mass-loss from low-mass stars, as the time-scale for their evolution off the main sequence is of $\gtrsim$$40 \, \mathrm{Myr}$. The growth of the relative abundance of AGB and supernova dust becomes slow and steady after some $\sim$$150 \, \mathrm{Myr}$.

Fig.~\ref{fig:AGBvsSNe} shows that the relative amount of astrating AGB/SN dust particles follows very closely the fraction of AGB/SN particles available in the ISM. Since the ratio has not reached a complete equilibrium by the end of our simulation, we cannot quantify the extent of the disparity between released and astrating grains. However, the general trend indicated by Fig.~\ref{fig:AGBvsSNe} is that even at enhanced or equivalent AGB particle production rates relative to supernovae, the latter contributes to a considerable proportion of star-formation events. At the end of the simulation, -- after more than $800\, \mathrm{Myr}$ of galactic evolution -- the AGB/SN ratio of accreted grains lies at around $\sim$0.8 or lower. As supernovae do not release more particles than giant stars on average after $200 \, \mathrm{Myr}$, the ratio on the order of unity does not result from an excess in dust production by supernovae. The details of our results are dependent on our stellar evolution assumptions \citep[see e.g.][on the evolution of stars with masses 7--$15 \, \mathrm{M_{\sun}}$]{Limongi_2024}, but slightly increasing or decreasing the AGB threshold mass should not qualitatively alter our results. For presolar grains, we could then expect at least around $50\%$ of dust grains to have originated from supernovae.\footnote{Bearing in mind that we do not distinguish between various chemical compositions of grains, but considering all types of grains our results are in agreement with the estimates of e.g. \citet{Hoppe_2022} that the supernova contribution to all presolar minerals is $>$$25\%$.}

Our calculations also show that AGB and SN dust particles end up forming a star between $\sim$1.5 -- $6 \, \mathrm{kpc}$ from the galactic centre, at a mean radial distance of $3.2 \, \mathrm{kpc}$ for the former and $3.4 \, \mathrm{kpc}$ for the latter. These distances correspond approximately to the exponential scale radius of the disc. Astration by these grains does not seem to occur on the outskirts of the galaxy, at distances equivalent to that of the Solar Neighbourhood ($\sim$$8.5 \, \mathrm{kpc}$). This might be due to the fact that the molecular gas depletion time-scales are on the order of $10\, \mathrm{Gyr}$ in the outskirts of the disc \citep[see e.g.][]{Leroy_2008}, and as a result, astration does not occur this far out for the limited duration of $<$$1 \, \mathrm{Gyr}$ of our simulation. The possibility of the Sun having radially migrated outwards from its birth location is supported by its enhanced metallicity with respect to stars of similar age and galactocentric location \citep{Wielen_1996}. The results from simulations of galactic dynamics by e.g. \citet{Minchev_2013, Tsujimoto_2020, Agertz_2021} suggest the Sun could have migrated from as far in as $r_{\text{GC}} \lesssim 5 \, \mathrm{kpc}$. However, the lack of star-forming grains on the galactic outskirts is, in our case, likely an artefact of our limited tracer particle resolution discussed in Section~\ref{sec:resolution}, along with not evolving the galaxy for long enough to capture the long time-scales present in the outer disc.

\subsection{Presolar grains}
\label{sec:presolar_grains}
\begin{figure}
    \centering
    \includegraphics[width=\columnwidth]{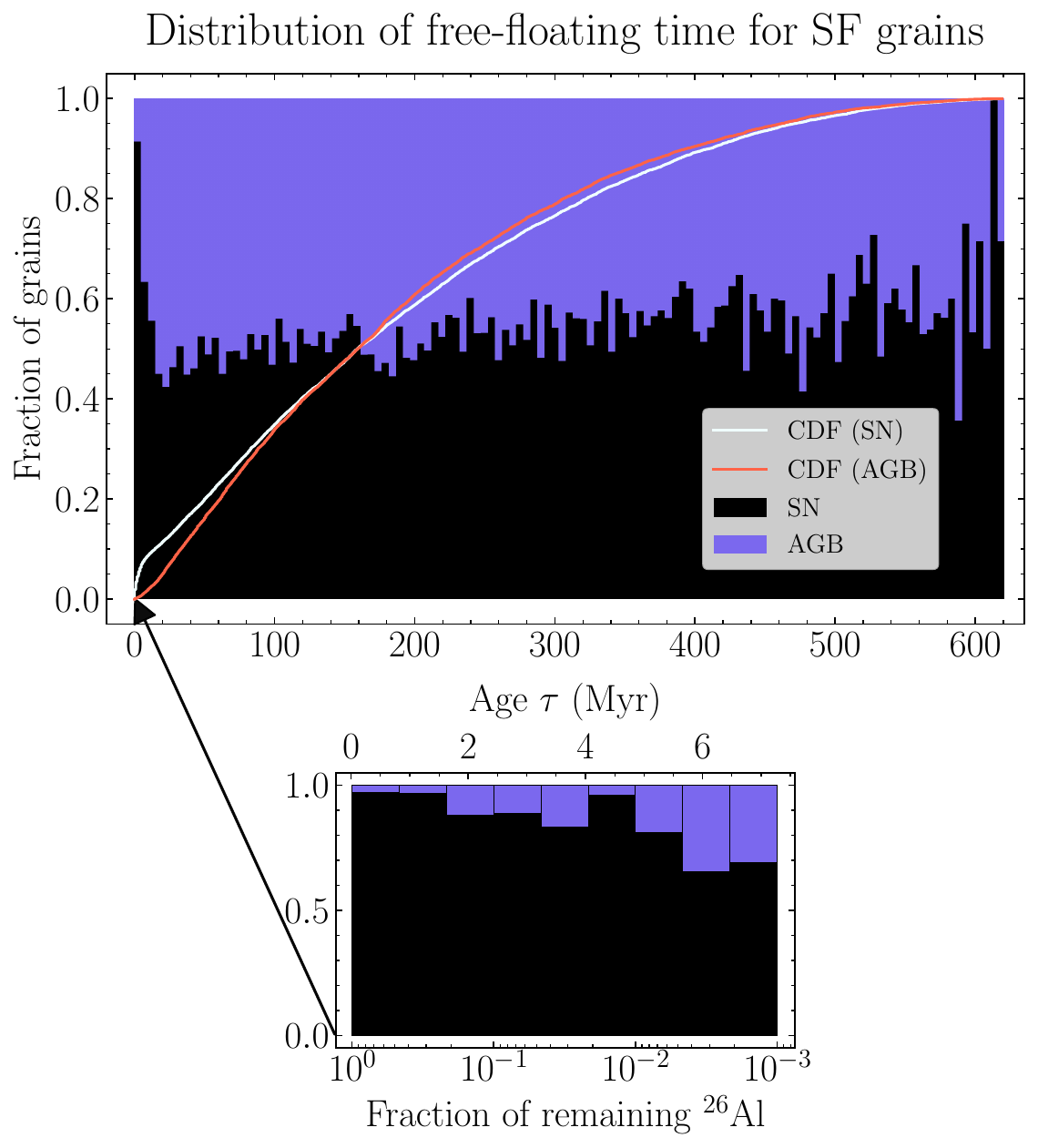}
    \caption{Distribution of the relative contribution from AGB and massive stars to star-forming grains, depending on their age $\tau$ binned in $5 \, \mathrm{Myr}$ intervals. Solid lines represent cumulative distributions for the filtered dataset of particles released after at least $200 \, \mathrm{Myr}$ -- which is why the CDF reaches 100\% at around $600 \, \mathrm{Myr}$. The lower panel shows the relative contribution from supernovae/their progenitors (black) and AGB stars (purple) to the remaining concentration of $^{26}$Al in the youngest astrating grains.}
    \label{fig:tau}
\end{figure}

We also consider the amount of time $\tau$ dust particles released from stars spend in the interstellar medium before undergoing astration. As explained in Section~\ref{sec:AGBvsSNe} and illustrated in Fig.~\ref{fig:AGBvsSNe}, the ratio of AGB to SN particles is skewed towards lower values at the beginning of the simulation due to the rapid evolution of stars with masses $\gtrsim$$8 \, \text{M}_{\sun}$. To mitigate this bias, for the next part we consider only the dust produced after at least $200 \, \mathrm{Myr}$ of galactic evolution. By investigating the time-scales between release and star formation displayed in Fig.~\ref{fig:tau}, we find that for most grain ages, the distribution is nearly identical between AGB and SN grains, with the exception of a few outliers for very old grains. It should be noted that due to the much smaller sample size of grains older than $600 \, \mathrm{Myr}$, on the order of a few to tens of particles, the distribution is highly sensitive to statistical fluctuations. In addition, these particles probe early times in the simulation when the AGB/SN ratio is low, as shown in Fig.~\ref{fig:AGBvsSNe}. The relative SN and AGB contributions to newly formed star particles by very old astrating particles should hence not necessarily be interpreted as being rooted in a physical explanation. The majority of the grains, however, undergo astration approximately $150 \, \mathrm{Myr}$ after their release -- in good agreement with presolar silicon carbide age estimates from e.g. \citet{Heck_2020}, who show that the majority of analysed grains from the Murchison meteorite spent less than $300 \, \mathrm{Myr}$ in the ISM prior to the formation of the Solar System. On the other hand, we find that particles that are rapidly accreted again after release are predominantly SN particles. As depicted in Fig.~\ref{fig:tau}, more than $80\%$ of grains younger than $5 \, \mathrm{Myr}$ have a supernova origin. The cumulative distribution also indicates that nearly $40\%$ of the SN and AGB particles are incorporated into a star again after less than $100 \, \mathrm{Myr}$.     

Due to the important role of $^{26}$Al in planet formation (see discussion in Section~\ref{sec:intro}), we quantify the relative amount of this isotope present in grains as they are reincorporated into a star, comparing the contribution of SN and AGB stars. The proportion of astrating grains that originate from AGB stars becomes significant only after about $10 \, \mathrm{Myr}$, yet by that time nearly no $^{26}$Al remains due to its radioactive decay. The heating of planetesimals by $^{26}$Al is therefore unlikely to hold its origins in the winds of AGB stars. Indeed, as evidenced by the zoom-in panel in Fig.~\ref{fig:tau}, more than $90\%$ of grains containing between $\sim$10--$100\%$ of their initial amount of $^{26}$Al came from supernovae. Down to concentrations on the order of less than a per cent, SN particles are still by far the main contributors. Our findings are in good agreement with e.g., \citet{Gounelle_2015} and \citet{Martinet_2022}, who find that massive stars\footnote{As a reminder, in our case SN particles represent both particles released by core--collapse supernovae and in the winds of massive stars (see Section~\ref{sec:tracers}).} are most likely to be responsible for the $^{26}$Al budget of the Solar System.  

\section{Discussion}
\label{sec:discussion}
Whilst our numerical approach to tracing the origin of the dust that forms stars has many implications for understanding presolar grains in meteorites, the post-processing technique has its limitations and our methodology is subject to some caveats, which shall be examined in this section.     

\subsection{Dust growth}
\label{sec:growth}
In addition to observational discrepancies on the importance of growth by accretion onto dust in the ISM \citep[e.g.][]{Nanni_2020, Galliano_2021}, the metallicity dependence of growth processes remains uncertain by orders of magnitude and hence their efficiency as well \citep[e.g.][and references therein]{Choban_2025}. None the less, accounting for the galactic dust content without growth is challenging \citep[e.g.][]{Dwek_1998, Inoue_2003, Pipino_2011} and the accretion of metals in the gas phase has widely been incorporated into dust evolution models in both galactic chemical evolution models and hydrodynamical simulations mentioned in Section~\ref{sec:intro}. 
For illustrative purposes, we introduce an additional growth term in the evolution of the grain size, derived from the accretion time-scales for silicates and carbonaceous dust of \citet{Hirashita_2014} (their equations (7)-(8)):
\begin{equation}
    \frac{{\text{d}}a}{{\text{d}}t} = \dot{a}_{\text{dest}} + \dot{a}_{\text{growth}} \, ,
    \label{eq:da_growth}
\end{equation}
where $\dot{a}_{\text{dest}}$ is the sputtering rate in equation~\ref{eq:sputtering_eq}, and
\begin{equation}
    \dot{a}_{\text{growth}} \approx 2 \times 10^{-13} \mathrm{\mu m \, yr^{-1}} \biggl(\frac{Z}{Z_{\sun}}\biggr) \biggl(\frac{n_{\text{H}}}{\mathrm{cm^{-3}}}\biggr)\biggl(\frac{T}{\mathrm{K}}\biggr)^{1/2} \biggl(\frac{S_{\text{acc}}}{0.3}\biggr)
    \label{eq:growth}
\end{equation}
is the rate of condensation from the gas phase. In equation~\ref{eq:growth}, $S_{\text{acc}}$ is the sticking efficiency, which we have assumed to be of the form $\exp(-T/T_{\text{acc}})$ with $T_{\text{acc}} = 100 \, \mathrm{K}$, supported by the calculations of \citet{Bossion_2024} for carbonaceous dust in gas at temperatures between $50 \, \mathrm{K}$ and $2250 \, \mathrm{K}$. Our approximation of the sticking coefficient is inexact but sufficient for our purposes, as the distinction between different chemical species of dust is not included in our model. The dependency of the sticking efficiency on the gas/grain temperature and chemical composition is more complex than in our formulation; however, the exponential assumption allows us to partially capture the temperature dependency while avoiding the assumption of $S_{\text{acc}} = 1$, which has been suggested to result in an overdepletion of metals by \citet{Zhukovska_2016}. It should also be noted that equation~\ref{eq:growth} is a crude approximation to probe the condensation of metals onto dust in the ISM, and as such, we decided not to include it in our nominal model.

\begin{figure}
    \centering
    \includegraphics[width=\columnwidth]{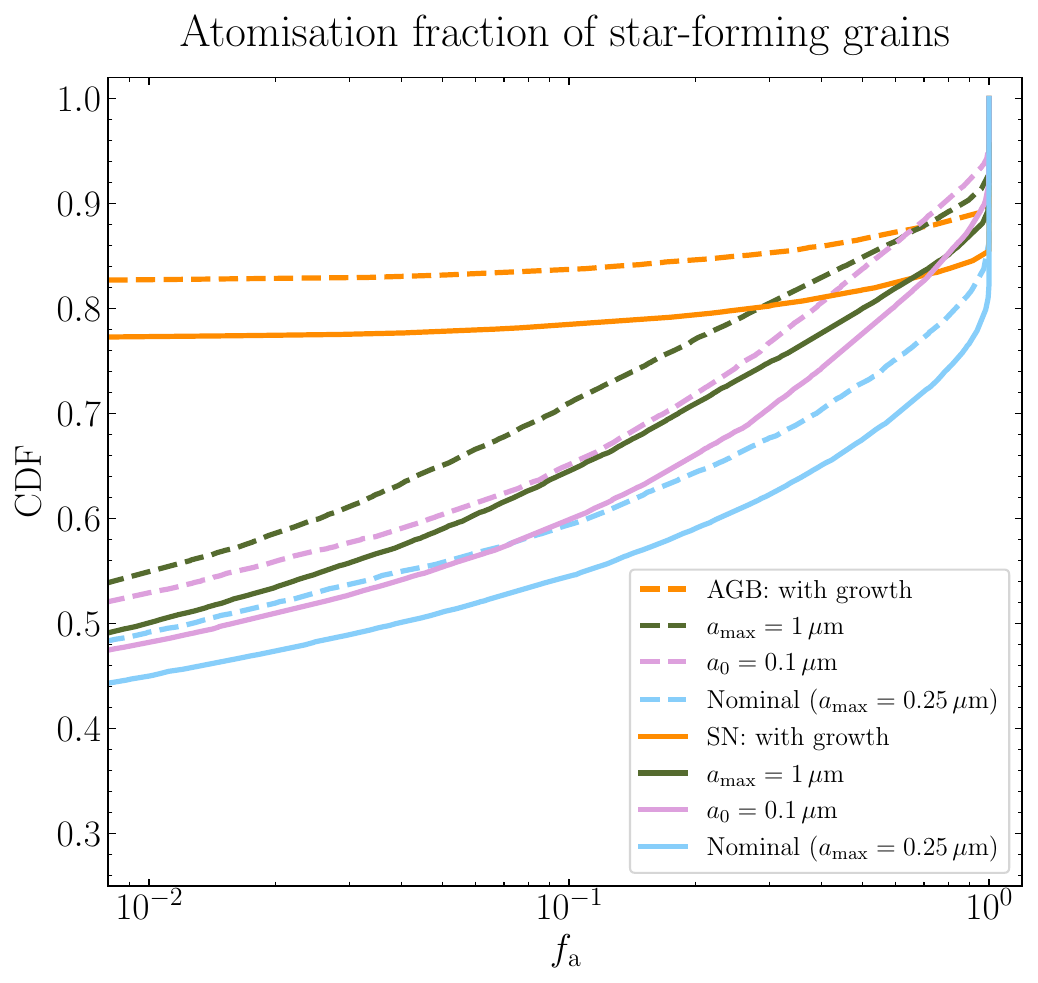}
    \caption{Cumulative distributions of mass lost due to thermal sputtering before undergoing astration for AGB (dashed) and SN (solid) dust particles for different initial grain size choices. Values for the model that includes dust growth, detailed in Section~\ref{sec:growth}, are shown in orange. The initial grain size distribution for the model with growth is identical to our nominal model -- a power-law distribution with $a_{\text{max}} = 0.25 \, \micron$ (see Section~\ref{sec:grain-size}). The pink lines represent the model in which we assume a constant initial grain size of $a_0 = 0.1 \, \mathrm{\micron}$ for all particles, the blue lines are our fiducial model and the green lines are for a power-law distribution with $a_{\text{max}} = 1 \, \micron$.}
    \label{fig:grain-size-test}
\end{figure}

When both the growth term of equation~\ref{eq:growth} and the thermal sputtering term of equation~\ref{eq:sputtering_eq} are considered in our model, the amount of particles (both for SN and AGB) with $f_{\text{a}} < 1\%$ nearly doubles relative to the model in which we consider only thermal sputtering. This is illustrated in orange in Fig.~\ref{fig:grain-size-test}. The significant decrease in their final degree of atomisation is due to their repeated and prolonged presence in the cold and dense ISM, where they grow in size and thus undergo astration having lost a smaller amount of their initial mass than in our nominal model (yet likely with a new chemical composition). The shape of the distribution is also flatter than in the model variations without growth, suggesting that most particles undergo astration with a very low degree of atomisation, while $\sim$$20\%$ and $\sim$$10\%$ of SN and AGB particles, respectively, have been completely atomised. These results imply that the introduction of growth in our model significantly reduces the number of small grains and biases the grain size distribution towards larger values. Our nominal model therefore potentially overestimates the extent to which dust grains have been atomised prior to becoming incorporated into a star again. In addition, the time-scales shown in Fig.~\ref{fig:atomisation_cats_tau} might be offset, as particles with long free-floating times in the ISM have the possibility to regrow by accretion. However, due to its effect on the size distribution and consequently also the final degree of atomisation of the grains, it is evidently preferable not to include dust growth in order to separate out the atomisation of dust grains in the ISM from dust growth. For our study of the relative contributions of AGB stars and supernovae to star- and planet-forming dust, adopting a model which focuses on the sputtering of dust with a well-defined initial size distribution offers clear benefits for the analysis.       

\subsection{Grain size distribution}
\label{sec:discussion-grain-size}
The choice of a power-law distribution for the initial grain size was motivated by empirical arguments. However, using a MRN distribution \citep{Mathis_1977} may not be applicable to all cases. The results of \citet{Sarangi_2015} suggest that the distribution of dust grains condensing in the ejecta of SNe type II-P differs from a power-law distribution and is skewed towards larger grains. \citet{Sluder_2018}, on the other hand, obtain a power-law distribution but with a steeper slope $n(a) \propto a^{-4.4}$ beyond a peak at $0.068 \, \micron$ for the supernova remnant of SN1987A. The discordant results between various works suggest a model sensitivity that prevents a singular choice of initial grain size distribution. 

An alternative presented in other analyses of dust evolution such as \citet{Esmerian_2022} is to disregard the initial grain size distribution entirely and assume a constant initial grain size for all grains of e.g. $a_0 = 0.1 \, \micron$. Upon doing so, we find no significant difference with our results, except for the fact that the mass-loss by thermal sputtering is reduced due to the lack of small grains, as is shown in Fig.~\ref{fig:grain-size-test}. We additionally investigate the effects of increasing the value of the upper bound of our power-law distribution $a_{\text{max}}$ to $1 \, \micron$, which resulted in an increase in dust survival by an additional $10\%$ (see Fig.~\ref{fig:grain-size-test}). \citet{Mathis_1977} noted that graphite grains have sizes ranging up to $1 \, \micron$, and as mentioned in Section~\ref{sec:grain-size} supernova dust may reach $\sim$$4 \, \micron$ \citep[see e.g.][]{Gall_2014}. The sharp cutoff at $0.25 \, \micron$ implies that our model might be slightly overestimating the efficiency of dust destruction due to thermal sputtering, yet our limited supernova bubble resolution likely compensates for this overestimation.  

The upper and lower bounds for the range of initial grain sizes spanned by our distribution may also seem odd if we consider that the dust condensed in the ejecta of core-collapse supernovae is unlikely to be of equivalent diameter and follow the exact same distribution as those condensed in the winds of giant stars. For the purposes of comparing the survival of SN to AGB grains through the ISM, however, using identical initial conditions has obvious advantages. We also note that our models do not include the distinction between grains of varying chemical compositions. \citet{Nozawa_2006} reported that different types of molecules exhibit diverse initial grain size distributions, which subsequently influence their destruction efficiency. Such considerations go beyond the scope of this paper and are left for future investigation.                   

\subsection{Dust-gas coupling}

Another important assumption in our work is that dust and gas are coupled. Temperature and collisional coupling is expected in regions of the interstellar medium where the density is $\gtrsim$$10^{4.5} \, \mathrm{cm^{-3}}$ \citep[see e.g.,][]{Goldsmith_2001, Hocuk_2017}. The results of \citet{Soliman_2024} also support the dynamics of small grains on the order of $0.1 \, \micron$ being closely determined by the motion of the gas, but suggest that the coupling is much less effective for larger grains with sizes of $\sim$$1 \, \micron$. The necessity of regarding dust and gas as dynamically distinct entities is further substantiated by \citet{Kirchschlager_2024a}, who establish that turbulence induces decoupling for large grains. Due to our choice of initial grain size distribution, which is strongly skewed toward small grains of $\lesssim$$0.1 \, \micron$, we regard the coupling of dust and gas as a reasonable assumption. Furthermore, decoupling effects occur at spatial scales well below the resolution level of our simulation, as described in Section~\ref{sec:SF}.  
On the other hand, we underline as a limitation to the interpretation of our results that the analysed simulation does not include magnetohydrodynamics. Explicitly incorporating the effects of dust at the scales resolved by \citet{Kirchschlager_2024a} in a galaxy simulation like ours would entail a significant increase in computational cost with, as mentioned in Section~\ref{sec:intro}, at only a modest improvement in physical realism. Nevertheless, their findings suggest that the effects of turbulence and magnetic fields may in fact inhibit dust destruction. It is therefore not impossible either that dust survivability exceeds our estimates.            

\subsection{Resolution and model}
\label{sec:resolution}
Our consideration of global galaxy formation models for the evolution of dust imposes a limit to the accuracy we are able to achieve. In this regard, several improvements could be made. With enhanced gas mass and spatial resolution, a more detailed picture of the evolution of dust in the ISM and the SNR could be obtained and allow linking the results of e.g. \citet{Kirchschlager_2024a} to the large scale processing of dust in a galactic setting. Ideally, the modelling of dust includes a wider range of disruptive and growth processes such as kinetic sputtering, shattering, photo-destruction\footnote{This is only relevant for small grains, see e.g. \citet{Nanni_2024}.}, as well as growth by accretion (see Section~\ref{sec:growth}) and coagulation \citep[for an extensive review on dust processing in the ISM and SN shocks see e.g.,][]{Schneider_2024}. These implementations require a more sophisticated physical representation of the galaxy, including radiative processes, ionising radiation, and magnetic fields. On top of these modifications, augmenting our tracer particle resolution by orders of magnitude would allow us to determine the mass-loss from stars in time in much greater detail. The results in e.g. Fig.~\ref{fig:tau} could then be refined, but we are optimistic that the qualitative picture remains. An even greater advancement would comprise resolving individual populations, star-by-star. Such progress is currently already being implemented in galactic scale models \citep[{\small INFERNO},][]{Andersson_2023}, as well as in smaller scale models, for instance, {\small STARFORGE} \citep{Grudic_2021}, which has been used to investigate the thermodynamics of GMCs by \citet{Soliman_2024}. Leveraging these advances and applying our approach could provide a means to bridge the scales, from the evolution of dust through the galaxy to zooming into protoplanetary discs.

\subsection{Extension to higher redshift}
The observational discrepancies mentioned in Section~\ref{sec:intro} motivate an investigation of dust evolution at higher redshift than $z=0$. From extragalactic observations we know that Milky Way-like progenitors were likely more compact, more gas-rich, and formed stars at a higher rate \citep[e.g.][]{Tacconi_2018}. In order to mimic a redshift $z\sim 1$ ($\sim$$8$ Gyr ago) Milky Way--mass galaxy, we run an additional simulation with an increased initial gas fraction of $f_{\text{g}} = 50\%$ while maintaining the other parameters identical. A similar set-up by \citet{VanDonkelaar_2022} showed that varying only the initial gas fraction can reproduce the star formation rate and stellar age--velocity dispersion relation of Milky Way--like galaxies in different redshift ranges. Due to its more violent evolution, with a star formation rate on the order of $25~{\rm M_\odot}$/yr (about 10 times higher than the $f_{\text{g}} = 10\%$ simulation), the considerable increase in computational load prevents us from running the simulation for more than $250 \, \mathrm{Myr}$. During that time, a steady-state for the relative production of AGB to SN dust particles (as in Fig.~\ref{fig:AGBvsSNe} for the $f_{\text{g}}=10\%$ simulation) is not reached, and therefore we cannot perform a similar analysis as for the lower gas fraction. 

None the less, by comparing the evolution of dust in the first $250 \, \mathrm{Myr}$ for both galaxies, we find that the thermal sputtering efficiency increases by about $\sim$$10\%$ for the high gas fraction galaxy. This is likely due to the more extreme physical conditions resulting from the higher gas fraction -- more gas available to form stars implies an augmented rate of supernova explosions and heating. Interestingly, supernova grains in our fiducial model are less severely eroded compared to AGB ones in the high-$z$ galaxy. We attribute this to the enhanced rapidity of reincorporation into a star by SN particles. However, we cannot conduct a detailed assessment of dust survival in a Milky Way--like galaxy at high redshift using this method. Along with the bias of AGB/SN particle production, \citet{VanDonkelaar_2022} point out that important aspects of galactic evolution, such as growth, as well as the gas accretion and merger history, are not taken into account when simulating an isolated disc with an enhanced initial gas fraction. In order to make rigorous predictions about the early formation of the Solar System in relation to presolar grains, it would be intriguing to investigate dust evolution in the context of a Milky Way progenitor at higher redshift in a fully cosmological context \citep[e.g.][]{Agertz_2021}. Bearing in mind all the aforementioned caveats, we therefore encourage subsequent research to explore this area in greater depth.

\section{Conclusions}
\label{sec:conclusions}
In this work, we have investigated the evolution of dust in post-processing of a simulation of a Milky Way--mass galaxy. We used Lagrangian tracer particles to follow the evolution of dust grains with initial sizes drawn from a power-law distribution. The erosion of the grains due to thermal sputtering in hot plasma was considered, and we performed a comparison between the destruction efficiency, as well as the time-scales between release and star formation for supernova and AGB dust. Our main findings can be outlined as follows.

\begin{itemize}
    \item The majority of tracers that represent stardust lose less than half of their initial mass by thermal sputtering, with a degree of atomisation corresponding to less than $1\%$ for nearly $35\%$ of the particles. Supernova grains are more severely sputtered than dust from AGB stars, on average, but only by an absolute difference of about $5$--$10\%$. After more than $800 \, \mathrm{Myr}$ of evolution, the mean degree of atomisation for SN dust particles is on the order of $30\%$. Due to our limited spatial and mass resolution, we do not capture the evolution of the supernova remnant in detail, and therefore likely underestimate the sputtering efficiency. However, even when considering some extreme physical conditions in the supernova remnant, we find that around $40\%$ of grains lose less than half of their initial mass due to thermal sputtering. 

     \item The release of grains from supernovae dominates over AGB stars at early times in the simulation due to the slower evolution of low-mass stars, but the relative production and abundance of AGB compared to SN dust particles rapidly increases until $\sim$$200 \, \mathrm{Myr}$, after which the growth stabilizes. The ratio of astrating AGB/SN particles fluctuates close to $0.8$ after several hundreds of $\mathrm{Myr}$ of galactic evolution, but is still slowly increasing. This ratio corresponds to a supernova contribution of approximately $55\%$ to presolar grains. In contrast, the data obtained from meteorites indicate a supernova contribution of $\sim$10--$30\%$ \citep[see][]{Hoppe_2022}. This difference could be explained by the fact that the AGB/SN ratio has not reached a complete equilibrium by the end of our simulation.  

     \item Our analysis also shows that, although astrating grains of most ages are roughly equally represented by AGB and SN dust, particles younger than $10 \, \mathrm{Myr}$ old primarily emanate from massive stars with masses $\gtrsim$$8 \, \text{M}_{\sun}$. In consequence, the latter are almost entirely responsible for the $^{26}$Al budget of the grains, as its rapid radioactive decay prevents dust incorporated into stars on longer time-scales from containing more than negligible traces of the isotope. The differentiation of planetesimals due to the decay of $^{26}$Al should then be the result of the contributions by massive stars to the presolar dust from which the protoplanetary disc forms, consistent with the results of \citet{Gounelle_2015}.
\end{itemize}

Our work combines the realism of evaluating dust evolution in the full complexity of a galactic setting with the reduced computational cost of a post-processing analysis. The implications of our findings with respect to presolar dust grains, and particularly short-lived radioactive isotopes such as $^{26}$Al, highlight the relevance of our method, which we believe can be applied to shed light on the life-cycle of dust and determine the stellar origin of presolar grains.   

\section*{Acknowledgements}
We thank Nick Gnedin for constructive feedback that helped us improve the structure of the manuscript. We also thank Anna-Maria Söderman for providing the simulation outputs, and Corentin Cadiou for insightful discussions and helpful contributions to the analysis.
OA acknowledges support from the Knut and Alice Wallenberg Foundation, the Swedish Research Council (grant 2019-04659), the Swedish National Space Agency (SNSA Dnr 2023-00164), and the LMK foundation. The computations and data storage were enabled by resources provided by LUNARC, The Centre for Scientific and Technical Computing at Lund University on allocations (LU 2024/2-57) and (LU 2024/12-33). AJ thanks the Carlsberg Foundation (Semper Ardens: Advance grant FIRSTATMO), the
Knut and Alice Wallenberg Foundation (Wallenberg Scholar Grant 2019.0442)
and the Göran Gustafsson Foundation.

%%%%%%%%%%%%%%%%%%%%%%%%%%%%%%%%%%%%%%%%%%%%%%%%%%
\section*{Data Availability}
The data underlying this article are available on request to the corresponding author.
 
%The inclusion of a Data Availability Statement is a requirement for articles published in MNRAS. Data Availability Statements provide a standardised format for readers to understand the availability of data underlying the research results described in the article. The statement may refer to original data generated in the course of the study or to third-party data analysed in the article. The statement should describe and provide means of access, where possible, by linking to the data or providing the required accession numbers for the relevant databases or DOIs.

%\newpage 


\bibliographystyle{mnras}
\bibliography{ref} % if your bibtex file is called example.bib

%\begin{thebibliography}{99}
%\end{thebibliography}

%%%%%%%%%%%%%%%%%%%%%%%%%%%%%%%%%%%%%%%%%%%%%%%%%%

%%%%%%%%%%%%%%%%% APPENDICES %%%%%%%%%%%%%%%%%%%%%

% \appendix

% \section{Some extra material}

%%%%%%%%%%%%%%%%%%%%%%%%%%%%%%%%%%%%%%%%%%%%%%%%%%

% Don't change these lines
\bsp	% typesetting comment
\label{lastpage}
\end{document}